\documentclass[draft]{agujournal2019}
\journalname{JGR Planets}
\usepackage[utf8]{inputenc}
\usepackage[english]{babel}
\graphicspath{{figures/}{./}}
\usepackage[authoryear]{natbib}
\usepackage{wasysym}

\renewcommand{\deg}{\ensuremath{^\circ}}
\def\deg{\ifmmode{^\circ}\else{$^\circ$}\fi}
\def\h2o{\ifmmode{{\rm H}_2{\rm O}}\else{H$_2$O}\fi}

\begin{linenomath*}
\end{linenomath*}

\begin{document}

\title{Stratigraphic and Isotopic Evolution of the Martian Polar Caps from Paleo-Climate Models.}

\authors{E. Vos$^1$, O. Aharonson$^1$ $^2$, N. Sch\"{o}rghofer$^2$, F. Forget$^3$, E. Millour$^3$,
L. Rossi$^4$, M. Vals$^4$, F. Montmessin$^4$} 

\affiliation{1}{Dept. of Earth \& Planetary Sciences, Weizmann Institute of Science, Rehovot, Israel.}
\affiliation{2}{Planetary Science Institute, Tucson, Arizona, USA.}
\affiliation{3}{Laboratoire de Météorologie Dynamique/IPSL, Sorbonne Université, ENS, PSL
Research University, Ecole Polytechnique, CNRS, Paris, France
}
\affiliation{4}{Laboratoire Atmosphères, Milieux, Observations Spatiales (LATMOS/CNRS), Paris, France.}

\begin{keypoints}
\item The NPLD growth rate strongly depends on perihelion position, in late stages accumulation can be of order a meter over a precession cycle. 
\item The isotopic stratigraphic signal in the polar caps experiences secular evolution, in addition to oscillations due to orbital elements.
\item Northern ice deposits are depleted in deuterium compared to the south, biasing the isotopic composition of the present-day atmosphere.
\end{keypoints}





\begin{abstract}
Exposed scarps images and ice-penetrating radar measurements in the North Polar Layered Deposits (NPLD) of Mars show alternating layers that provide an archive of past climate oscillations, that are thought to be linked to orbital variations, akin to Milankovitch cycles on Earth. We use the Laboratoire de M\'et\'eorologie Dynamique (LMD) Martian Global Climate Model (GCM) to study paleoclimate states to enable a better interpretation of the NPLD physical and chemical stratigraphy. When a tropical ice reservoir is present, water vapor transport from the tropics to the poles at low obliquity is modulated by the intensity of summer. 
At times of low and relatively constant obliquity, the flux still varies due to other orbital elements, promoting polar layer formation. Ice migrates from the tropics towards the poles in two stages. First, when surface ice is present in the tropics, and second, when the equatorial deposit is exhausted, from ice that was previously deposited in mid-high latitudes. The polar accumulation rate is significantly higher when tropical ice is available, forming thicker layers per orbital cycle. However, the majority of the NPLD is sourced from ice that temporary resided in the mid-high latitudes and the layers become thinner as the source location moves poleward. The migration stages imprint different D/H ratios in different sections in the PLDs. The NPLD is isotopically depleted compared to the SPLD in all simulations. Thus we predict the D/H ratio of the atmosphere in contact with NPLD upper layers is biased relative to the average global ice reservoirs.
\end{abstract}

\section{Plain Language Summary}

In this work we run simulations of a Global Climate Model for Mars with a broad range of orbital elements, and ice initially placed in the tropics, we calculate the growth rate and hydrogen isotopic composition of the polar caps. These simulations help to understand the migration of ice from the tropical region to the polar caps, as believed to have occurred in the recent past. Ice transport to the poles occurs at two stages. The first is when ice is present in the tropics. The second is after the tropical reservoir has been exhausted, and the source of vapor that reaches the poles is from ice accumulated in mid-high latitudes during the first stage. The polar caps growth rate during the first stage is larger and results in thicker layers. We find that both physical and chemical records are expected in the polar caps, controlled by the orbital elements and the surface ice distribution. The chemical record also depends on the source enrichment. These results should help to interpret ice records in order to decode past climate variations, and suggest the current hydrogen isotopic composition of the atmosphere is not representative of the total ice reservoirs on Mars.

\section{Introduction}
Mars harbors ice deposits in several forms, the most conspicuous are the North Polar Layered Deposits \citep{byrne2009}. Spiraling troughs cut into the polar cap and expose a stratigraphy of alternating layers. Evidence that layers are generally flat and sub-parallel is found in orbital images of extended troughs and radar data  \citep[][and many others]{fenton2000,fishbaugh2006,milkovich2005,smith2016}. There are also visible layers in the South Polar Layered Deposits (SPLD) \citep[e.g.][]{plaut2007}, but because of their older age, a permanent CO$_2$ cover and the limited extent of exposed troughs, the stratigraphy of the SPLD has been studied in less detail \citep{becerra2019,milkovich2008}. Analysis performed on the upper NPLD stratigraphy found a dominant wavelengths of $\sim$30~m \citep{laskar2002,hvidberg2012,milkovich2005}, and 1.6~m \citep{perron2009,limaye2012,fishbaugh2010}. Different characteristic wavelengths are often taken to indicate variations in polar ice/dust fluxes, which may result from changing insolation regimes or ice source availability. SHARAD data show multiple reflections \citep{smith2016} that are caused by a change in the dielectric permittivity, implying transitions between ice and dust layers. The dust-rich layers are formed either at low levels of ice accumulation or by sublimation leaving lag layers \citep{hvidberg2012} with a possible unconformity in the stratigraphy \citep[e.g.][]{smith2016,vos2019}. Thus it is thought that the polar cap stratigraphy offers a record of past climate conditions,  expected to respond to the known large variations in Mars's orbital elements \citep{levrard2004,smith2020,laskar2002,byrne2009,vos2019,levrard2007,cutts1982,becerra2017}. However, establishing a connection between the layering and the orbital elements has proven difficult, due to the complex nature of the stratigraphic record, as well as  modeling uncertainties that include the role of dust, ice distribution, and other physical properties  \citep{madeleine2009}. 

\citet{laskar2004} calculated the orbital elements over the last 21~Myr.
During this time, the obliquity has varied between 15\deg\ and 45\deg, and the eccentricity between 0 and 0.13. Approximately 4.5 Myr ago, the mean obliquity dropped from $\sim$35\deg{} to $\sim$25\deg, resulting in lower mean polar insolation during the latter part  of the Martian history until the present.
It is during this epoch that equatorial ice is thought to have migrated to the poles \citep{levrard2007}.
Flux of ice to the poles is correlated with the orbital elements of the planet, in particular with obliquity which has a dominant period of 120~kyr. The length and intensity of the seasons are also governed by the eccentricity and perihelion precession, which have periods of 100~kyr and 51~kyr, respectively. 
Previous modeling work using GCMs \citep{levrard2004,levrard2007,emmett2020} demonstrated that at low obliquity, when a tropical ice source is available, net transport of ice is from the equatorial region to the poles. Net polar ice gain occurs when obliquity is below a threshold of $\sim$35\deg{} for a circular orbit, but the precise value depends on other orbital elements, such as the perihelion position, and on assumed model parameters such as dust and ice properties \citep{madeleine2009}. 


On Earth, The isotopic signal in ice cores from Antarctica and Greenland have revealed past climate oscillations \citep{alley2006}. On Mars, \citet{vos2019} used a 3-box model to predict the historic D/H record in the polar deposits. Mars lacks an ocean to buffer the atmospheric isotopic ratio as on Earth, thus the isotopic composition of the present atmosphere is driven by the evaporative fluxes of the caps. Observations from orbiters and ground-based telescopes measured the atmospheric D/H ratio to be in the range of 1-8 $\times$ Vienna Standard Mean Ocean Water (VSMOW) \citep{jakosky2018,villanueva2015,krasnopolsky2015,webster2013}, with an average of 4.6$\pm$0.7 \citep{krasnopolsky2015}. The atmospheric values provide evidence for both net water loss. 
in the long-term evolution of Mars \citep{alsaeed2019,villanueva2015}, as well as seasonal, cloud physics, and geographic source effects \citep{montmessin2005}.

Here, we provide quantitative results from the LMD-MGCM simulations that shed light on the stratigraphic and D/H isotopic evolution of the Martian polar caps, in particular the NPLD, the simulations are initialized with a tropical ice reservoir that progressively migrates and accumulates at the polar caps. We explore the range of orbital configurations experienced by Mars over the past 21 Myr to study the relation among surface ice distribution, atmospheric humidity and polar flux. We also investigate the migration stages of ice from the tropics to the poles that is believed to occur in the last $\sim$~4.5 Myr \citep{levrard2004,laskar2002,hvidberg2012,vos2019,cutts1982,smith2020,smith2016} and present hypotheses to explain differences in layer thicknesses in the NPLD. Lastly, we examine if the current atmospheric isotopic ratio is representative of the global average in the ice.

\section{Climate Model Description} 
\label{s:model}
The results reported in this paper are based on simulations of the Mars LMD-GCM with the full water cycle that includes treatment of surface ice, atmospheric vapor and ice clouds and has been described in detail previously \citep{forget1999,montmessin2004,madeleine2009,madeleine2011}.Therefore, we limit our model description to presentations of the most important assumptions and model configurations.

The time and space evolution of water vapor for the present configuration was compared and validated with the MGS-TES atmospheric humidity data \citep{smith2004,montmessin2004}. In order to track the D/H ratio we use the HDO tracer described in \citet{rossi2021} in both the vapor and ice phases, in addition to the dust and water tracers \citep{montmessin2005}. Temperature-dependent isotopic fractionation of water occurs at condensation either in clouds or directly onto the surface. The equilibrium solid-vapor fractionation coefficient ($\alpha$) assumed is based on lab measurements \citep{merlivat1967} at temperatures above 233~K. and was later tested extended to a wider range of temperatures as low as 194K \citep{lamb2017} with good agreement. 
The fractionation is calculated at each atmospheric grid cell, and at each time step, allows for geographic and seasonal variations that were not present in the simple 3-box model of \citet{vos2019}. We follow the "semi-interactive" dust scheme described in \citet{madeleine2011}, in which dust is lifted and vertically mixed according to the simulated wind dynamics that the model predicts.  The column-integrated opacity is scaled to match observations \citep{millour2018}. The dust is radiatively active, altering the atmospheric energy budget. We also tested a case with a constant dust level ($\mathrm{\tau}_\mathrm{dust}$=0.2, the amount of dust in the atmosphere is given by its optical depth, see Eq. 1 of \citet{madeleine2011}) to verify that the results are robust to this assumption.
The radiative effect of ice clouds is not taken into account, which can induce moderate temperature changes in moist atmospheres \citep{madeleine2014} of up to 10\deg{} K \citep{madeleine2012}. Additional work is required to quantify this effect when simulating past conditions.
The simulations reported here have a horizontal resolution of 64 by 64, corresponding to 2.8125\deg in latitude, and 5.625\deg in longitude. We increased the latitude resolution in comparison to previous studies \citep{madeleine2011,levrard2007} to better sample the polar region. There are 29 unequal vertical layers that span the atmosphere up to 100 km, and 18 unequal subsurface layers that mimic the penetration of heat in the ground up to 18.5 m. We ran the GCM for paleo-orbital element combinations \citep{laskar2004}: obliquity between 15\deg{} and 45\deg{}, eccentricity between 0 and 0.13, and different longitudes of
perihelion ($L_p$) spanning the full range. We ran $\sim$~60 different orbital combinations that cover the parameter space mentioned above, a large increase from previous work. The simulations were performed with an assumed past ice distribution in which surface ice is present in the tropics at locations on the eastern flanks of the Tharsis rise (with an area of $\sim10^6 km^2$) where accumulation is predicted at high obliquity \citep{forget2006}. The presence of such a past tropical surface ice reservoir is supported by geologic evidence \citep{head2005}.
We tested thick and thin tropical ice layers. In the thick case, the tropical ice layer has a thickness of $\sim$60~m Polar Equivalent Layer (PEL), defined as a layer of the same volume, spread evenly from 81\deg{} to 90\deg{} latitude. This case is designed to reach steady state, to investigate the fluxes and D/H ratio of ice reaching the polar cap with a persistent tropical source. In the thin test case, we initialize the model with a $\sim$0.45~m PEL tropical ice layer, in order to investigate the different stages of the evolution of ice migration as the finite initial source is exhausted.

\section{Results}
The results reported here summarize the dependence of expected accumulation rates and D/H ratio on orbital configuration and ice distribution at various assumed boundary conditions.

\subsection{Thick tropical ice reservoir}
\begin{figure}[tb!]
\centering
\includegraphics[width=\linewidth]{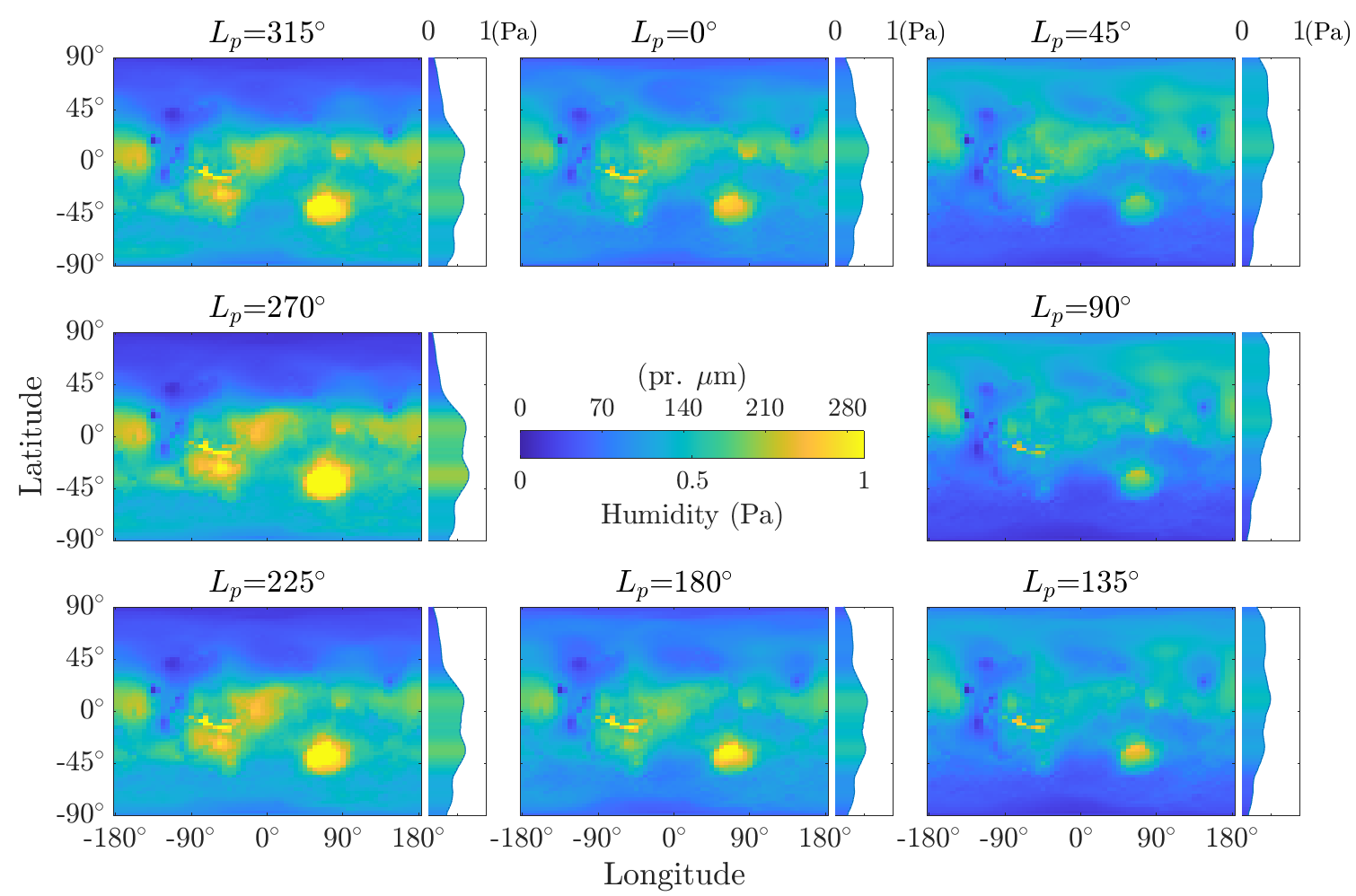}
\caption{Global maps of simulated surface annual average humidity for an initially thick ice distribution, zonal means are shown on the right of each panel. Orbital parameters are: obliquity = 25\deg{}, eccentricity = 0.093, and different values of $L_p$ are shown in each panel, sampling a full precession cycle. Units of integrated vapor column abundance are also shown. The eight panels sample a full precession cycle. The humidity is significantly higher than today's due to the tropical source. A short and intense summer in one hemisphere leads to increased transport of water vapor to the polar regions of that hemisphere.}
    \label{f:humidity}
\end{figure}

We begin with GCM simulations performed at various past orbital conditions with an initial thick tropical ice reservoir. Atmospheric water vapor was tracked in a suite of simulations with various orbital parameters. The humidity regimes are shown in Figure~\ref{f:humidity} for an obliquity of 25\deg{}, eccentricity of 0.093, as at present, and eight different values of $L_p$, spanning a full precession cycle. The humidity is significantly higher than today's, owing to the vapor source in the equatorial region. There are annual mean local enhancements of $\sim$280~pr.~$\mu$m (precipitable microns, vertically integrated in the atmosphere), whereas today, the peak is $\sim$50~pr.~$\mu$m above the North Polar Cap at the northern summer maximum with a much lower annual mean \citep{smith2004}. When perihelion is aligned with the spring or autumn equinox ($L_p$=0\deg{} and 180\deg{} respectively) the annual mean water vapor distribution is roughly symmetric. More intense, shorter summers in one hemisphere (e.g., $L_p$=90\deg{} in the North) drive more water vapor to migrate poleward in that hemisphere, due to the increased atmospheric humidity at warmer temperatures. These differences in meridional vapor flux with $L_p$ lead to variations in polar accumulation rate, as will be shown below. 

\begin{figure}[tb!]
    \centering
    \includegraphics[width=\linewidth]{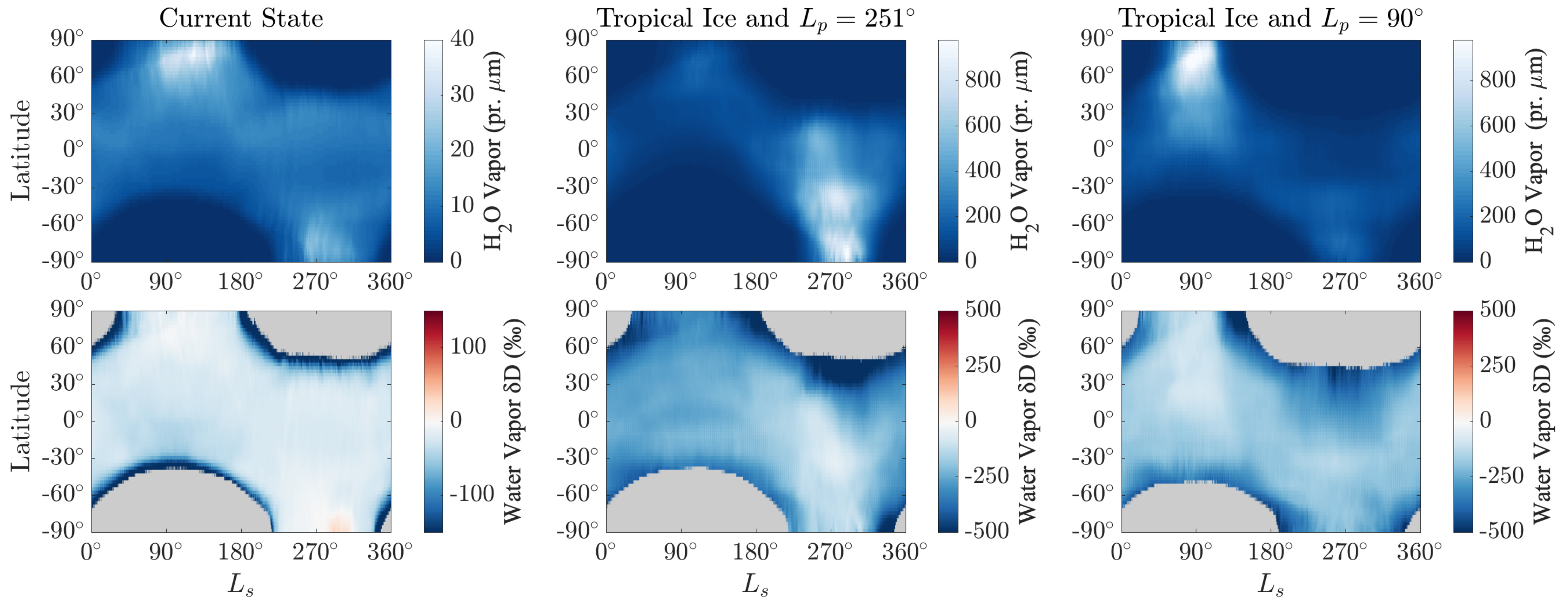}
    \caption{Zonal mean seasonal evolution of atmospheric water vapor and its D/H ratio for three different states: Current orbit and ice distribution (left column), current orbit and a thick tropical ice distribution (middle column), and an orbit with $L_P=90\deg$ and a thick tropical ice distribution (right column). With a tropical ice reservoir, the atmospheric humidity and the D/H ratio are higher in the hemisphere at which summer aligns with perihelion, i.e. it is short and intense. Note the larger scale of the humidity and $\delta{}D$ when a tropical source is present.}
    \label{f:Ls_lat_today}
\end{figure}

The zonal mean seasonal evolution of water vapor and atmospheric water vapor D/H relative to the initial value of the source, $(D/H)_0$ defined as $\delta$D$=\left(\frac{(D/H)_{\ }}{(D/H)_0}-1 \right)\times{}1000\permil$. are shown in Figure~\ref{f:Ls_lat_today},  comparing cases with difference orbital state and ice distribution. Atmospheric humidity is significantly higher when a tropical ice reservoir is available, and peaks near mid-summer in each pole ($L_s$=90\deg{} in the North and 270\deg{} in the south). In general, the atmospheric vapor is isotopically depleted because the HDO is either preferentially in ice clouds or in surface ice that sedimented at an earlier stage. This phenomenon is amplified when a tropical reservoir is present due to the high humidity. Therefore, the atmospheric water vapor D/H ratio is higher in the poles when the summer is short and intense, because there is a large amount of water vapor, and thus lower fractionation.

\begin{figure}
\centering
\includegraphics[width=\linewidth]{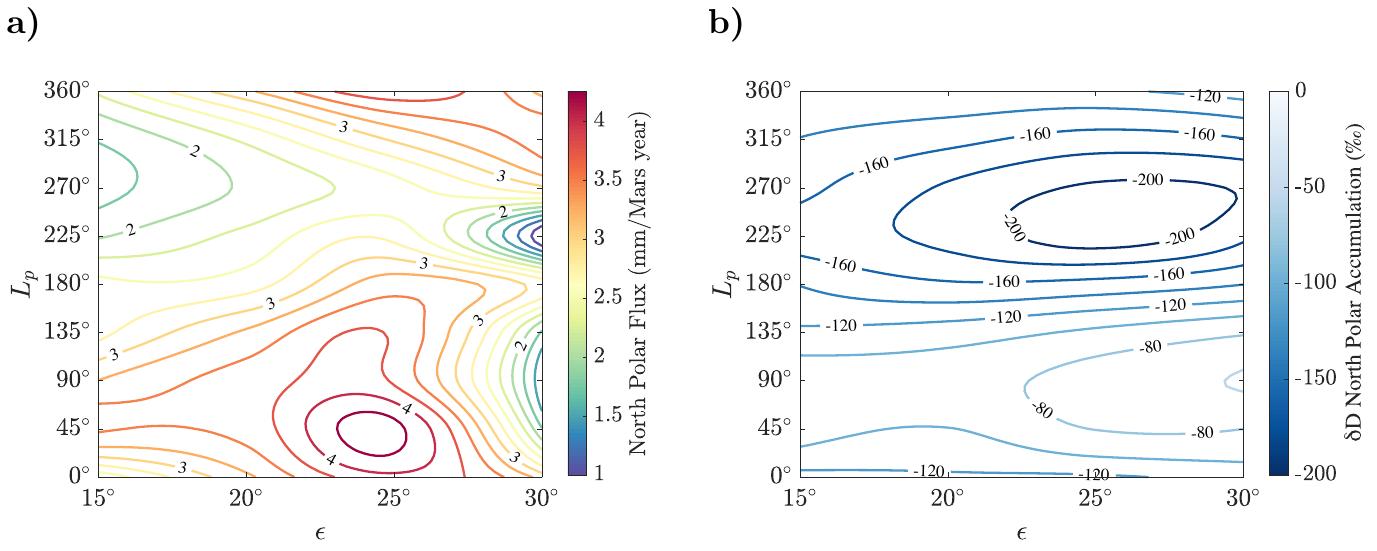}
\caption{{\bf a)} Simulation results of the North Polar ice flux as function of obliquity noted as $\epsilon$ and $L_p$ for eccentricity of 0.093 with a thick tropical reservoir. A short and more intense summer leads to higher accumulation rate up to obliquity $\sim$27, beyond which it decreases rapidly.
{\bf b)} North polar ice accumulation D/H ratio for the same orbital configuration and ice distribution as in a. A short, intense summer leads to higher D/H ratio. The isotopic layering is mostly determined by $L_p$. \label{f:Polar_accu}}.
\end{figure}

In addition to the atmospheric humidity described above, the evolution of surface ice was also tracked. 
Following past work \citep{levrard2007,emmett2020,levrard2004,madeleine2009}, we track the evolution of surface ice at past orbital parameters. Here, we significantly extend the range of conditions studied, use more realistic ice distributions that are based on high obliquity GCM simulations and geologic evidence \citep{forget2006,head2005}, and provide interpretations based on the humidity patterns above, to derive a relation between polar cap growth rate and orbital elements. The NPLD (defined here as the area polewards of 81\deg{}) growth rate as a function of obliquity and $L_p$ is shown in Figure~\ref{f:Polar_accu}a, for present-day eccentricity. In addition to obliquity controlling the annual mean insolation as a function of latitude, and hence the sign and magnitude of polar accumulation, there is also a strong dependence on $L_p$. This phase controls the length and intensity of the high-humidity summer season, with a difference of a factor of $\sim$2 in accumulation rate between a prolonged mild northern summer to a short intense one. Higher eccentricity simulations show this effect is amplified and the NPLD growth can reach 6~mm/Mars year. 
At obliquities higher than 25\deg{} the pattern changes, the dependence becomes more complicated, and the mean accumulation rate no longer shows a single peak in $L_p$ over a precession cycle. This occurs because with a tropical humidity source at moderate obliquities there is continuous polar accumulation throughout the year. At obliquities $\sim$30\deg{} and above, polar ice begins to sublimate during the summer season; at even higher obliquities, the NPLD flux is at a state of net loss. Overall, the simulations show the growth rate of the north polar cap is 1-6~mm/Mars year, while loss rates can reach 100~mm/Mars year.  \citet{levrard2007} found similar loss rates.
The results in Figure ~\ref{f:Polar_accu}a imply that the stratigraphy in the cap is controlled not just by obliquity cycles, but also strongly, by precession cycles.

To study the hydrogen isotopic signal expected in the Martian polar cap, we analyze the results of the same simulations presented above. We calculate the mean D/H ratio of ice that condensed in the polar region and plot it in Figure~\ref{f:Polar_accu}b.
The contours show D/H values in $\delta$D.
The hydrogen isotopic anomaly depends on the orbital elements, and in particular $L_p$, with a difference of more than 100\permil{} over a precession cycle. A short, intense summer leads to higher D/H ratio of the accumulating ice ($L_p$ of 90\deg{} for the NPLD and $L_p$ of 270\deg{} for the SPLD, not shown in the figure). Notably, the polar $\delta$D value is negative for all cases. This occurs because as vapor transports from the tropics poleward, the water molecules first deposit in the intermediate latitudes and preferentially condense out the heavy isotope. The atmosphere becomes depleted in the HDO and therefore water reaching the pole is isotopically light.  Under the assumption that the majority of the tropical source ultimately migrates to the polar cap, the depletion in HDO of the early deposits would be compensated by enrichment of subsequent deposits sourced from the intermediate latitudes. These results support the notion that an isotopic signal should be expressed in the vertical structure of the polar caps \citep{vos2019}. 

\begin{figure}[tb!]
    \centering
    \includegraphics[width=\linewidth]{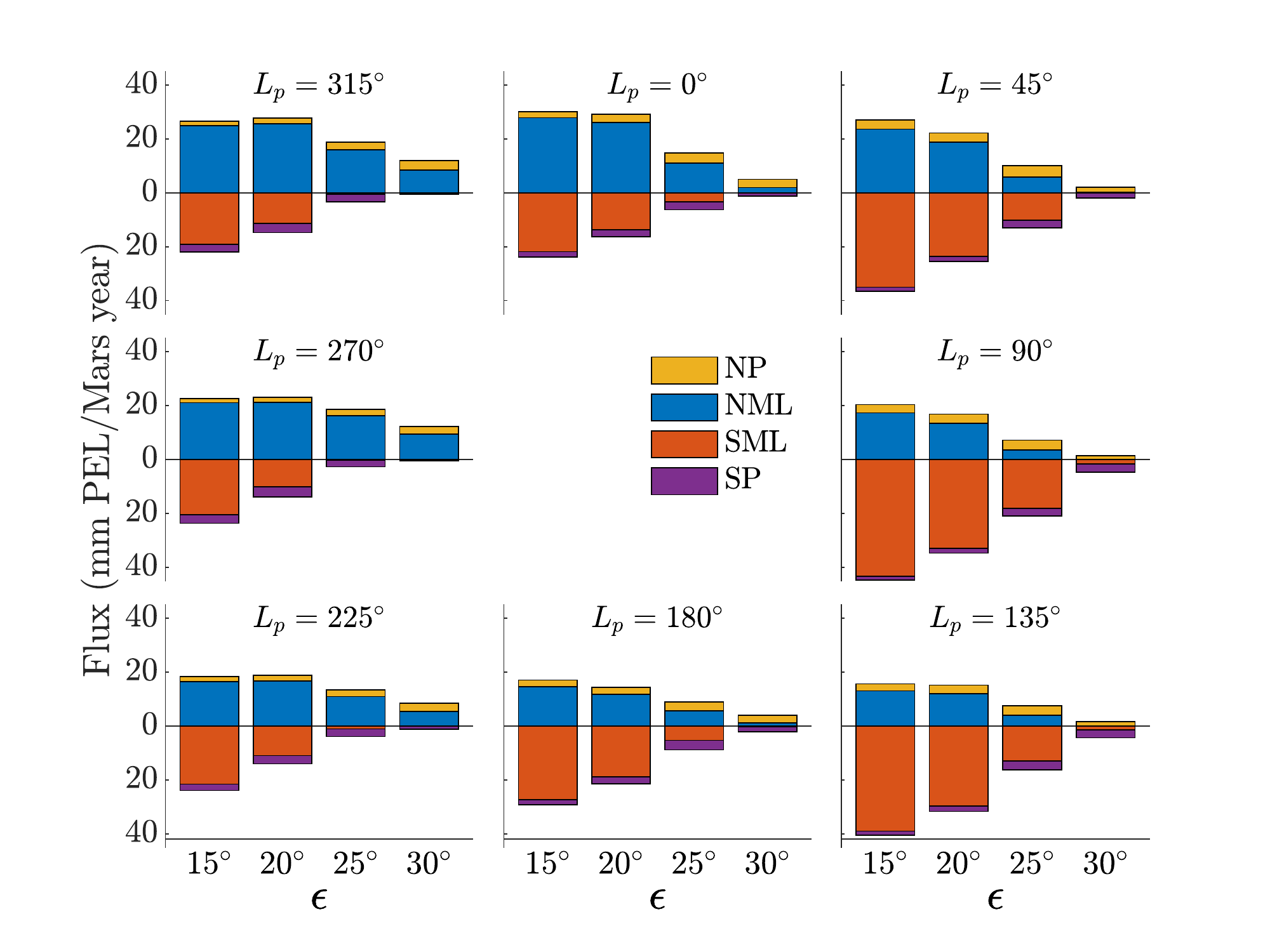}
    \caption{Flux to the North Polar Cap (NP), North Mid-high Latitude (NML), South Mid-high Latitude (SML), and South Polar Cap (SP) as function of obliquity and $L_p$ in units of Polar Equivalent Layer (PEL). The sum of all the components corresponds to the flux from the initial reservoir placed in the equatorial region. Lower obliquity results in a higher loss from the tropics, and a greater fraction of the ice being trapped at mid-high latitudes. North-south asymmetries in ice accumulation are seen.}
    \label{f:bar_plot}
\end{figure}

In addition to the dramatic increase in humidity and polar ice accumulation rate that occurs when surface ice is present in the tropics, significant ice condensation also occurs at intermediate latitudes during these times \citep{levrard2004}. To study the transport of ice from the tropics to these latitudes, we integrated the accumulation rate over five broad bands: the equatorial region (here defined as 25\deg{}S to 25\deg{}N, including the tropical source), the mid-high latitudes in each hemisphere (25\deg{} to 81\deg{}), where ice temporarily resides during its poleward migration, and the polar regions in each hemisphere (81\deg{} to 90\deg{}). The results are summarized in Figure~\ref{f:bar_plot}. The amount of ice that reaches each pole is 1-6~mm/Mars year, only a fraction of the 10's of mm/Mars year lost from the tropics, with the remainder being trapped at mid-high latitudes. While the tropical loss rate is primarily determined by obliquity, the fluxes at polar and mid-high latitudes are also controlled by the seasonal asymmetry ($L_p$). For example, at obliquity $30$\deg{}, the mid-high northern latitude fluxes reduce from $>10$~mm~PEL/Mars~year to nearly zero over a precession cycle. Importantly, as can be seen in Figure~\ref{f:bar_plot}, only a small fraction of the ice that was lost from the tropical reservoir reaches the polar caps when ice is still available in the tropics; the majority is first deposited in the mid-high latitudes.
There are north-south differences in ice accumulation at opposite $L_p$'s, due to the asymmetric distribution of the initial ice deposits on the Tharsis rise \citep{forget2006,head2005}, although such asymmetries may also be caused by topographically induced dynamics as shown by \citet{Richardson2002}. 
 
\subsection{Thin tropical ice reservoir}
\begin{figure}
\centering
\includegraphics[width=\linewidth]{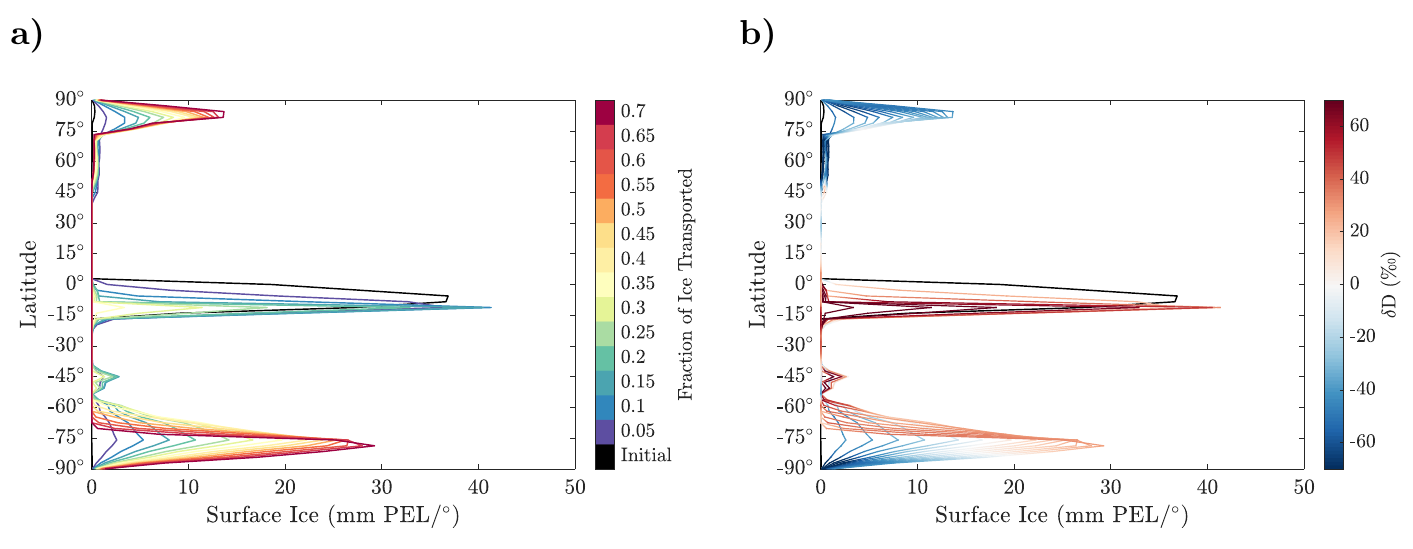}
\caption{{\bf a)} Surface ice evolution as function of latitude and ice fraction transported, for our nominal orbital configuration and $\sim$0.45~m PEL of ice placed in the equatorial region on Tharsis (see Section 3).
{\bf b)} Surface ice in PEL and $\delta$D of the layers as function of latitude for the same run as in a. The contours are plotted in intervals of 0.05 in fraction of ice transported. The colors represent the D/H ratio of the mean value of the layer that was formed. At each time the location of the D/H enriched ice is the equator-ward most latitude at which ice condenses, leaving the atmospheric composition depleted of the heavy isotope, so ice that condenses poleward is isotopically light. \label{f:surf_ice}}.
\end{figure}

In order to simulate the stages of poleward migration of ice from the tropics, we considered an initial ice distribution with a smaller amount of $\sim$0.45~m PEL of ice (thin case, see Section 3), placed in the equatorial region. We use this approach as an approximation, 
as the stages of the migration of this limited deposit over the simulation timescale reflects those of a much larger deposit, migrating over a much longer time. This approximation neglects effects that occur on millennial timescales such as the change in thermal inertia of the ground with increasing ice content, but it does capture the interplay between the reservoirs as they evolve.
The reservoirs evolve smoothly from year to year, and over the timescale of the simulation (100's of Mars years) they experience significant change.  
Figure~\ref{f:surf_ice}a shows this evolution for a reference orbital element combination, with an obliquity of 25\deg{}, eccentricity of 0.093 and $L_p$ of 90\deg. The simulation shown was run for 105 Mars years, but it is plotted as a function of the fraction of ice transported and not time, keeping with the idea that the evolutionary stages are maintained on longer timescales.
Indeed, we obtained similar results as a function of fraction of ice transported when we increased the amount of initial ice tenfold (see Figure~1 in the Supplementary Material), supporting our approach to simulate the migration stages in the thin ice case. 
At first, as long as the tropical reservoir is present, ice migrates to latitudes higher than $\pm$45\deg, but only a small portion of the ice that sublimated from the tropics reaches the polar caps. For most of the orbital element combinations we examined, ice also shifts from the tropics to the western flanks of Argyre Planitia. When ice is no longer available in the tropics, the planet's mean humidity drops and ice migrates from these quasi-stable regions towards higher latitudes, poleward of about $\pm$55\deg. As the simulation advances and more ice migrates, the location of the source (the equator-most ice still available) migrates poleward. At the last stage, when the boundary has migrated to $\sim\pm$75\deg, the ice accumulating on the caps is sourced only from these high latitude locations (either in the same hemisphere, or the opposite one).
Thus, the upper layers of the cap are expected to represent a deposit that migrated from high latitudes (polewards of 70\deg{}), and not directly from the tropical reservoir. Overall, the majority of the ice that comprises the polar caps did not migrate directly, but rather temporarily resided in the mid-high (25\deg{}-81\deg{}) latitudes before reaching the cap. 

The migration stages mentioned above also affect the D/H ratio of the condensed ice. Figure~\ref{f:surf_ice}b also plots the surface ice evolution like \ref{f:surf_ice}a, but here the colors represent the D/H ratio in units of $\delta$D. The ratio shown represents the mean value of the layer formed in the same intervals plotted in Figure~\ref{f:surf_ice}a. The arrows indicate the directions of growth/retreat as time progresses. The simulations show that at any given instant, the most equatorial latitudes at which ice condenses are also most enriched in the heavy isotope.
The remaining water in the atmosphere is relatively depleted in HDO, so ice that condenses poleward of this latitude is likewise depleted. Thus, the D/H ratio of growing ice on the polar cap depends not only on orbital parameters, but also on the source location and composition after the tropical reservoir is exhausted.

\begin{figure}[tb!]
    \centering
    \includegraphics[width=\linewidth]{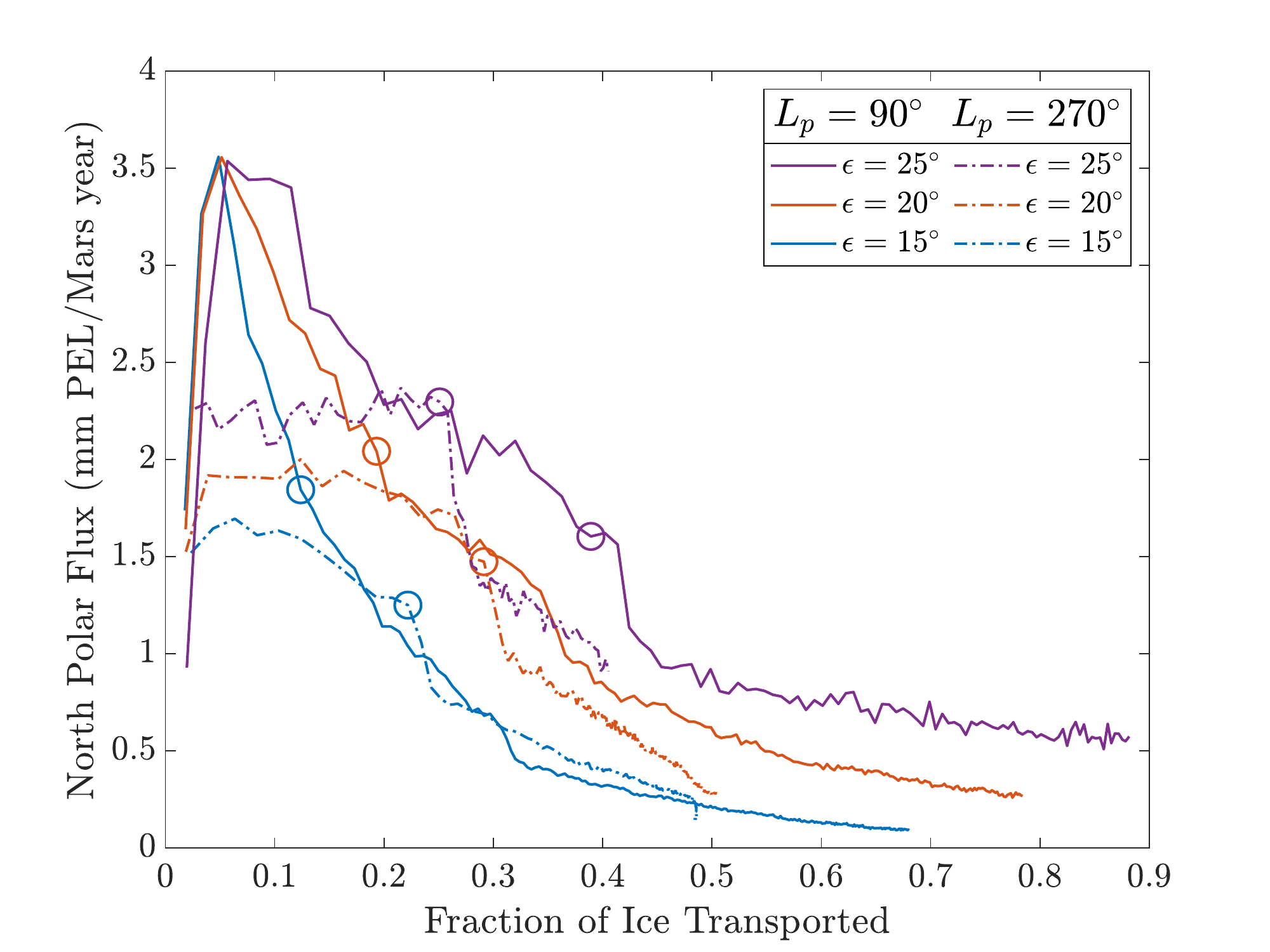}
    \caption{North polar flux as function of the fraction of surface ice that reached the polar caps. The circle markers represent the end of the first stage, when the tropical reservoir is exhausted (defined as 98\% removed). The fraction of ice that reaches the poles during the first stage is mostly determined by obliquity. At the second stage, when no more ice is available in the tropics, there is a dramatic decrease in flux, as the belt of the source ice moves poleward.}
    \label{f:flux_trans}
\end{figure}

The stages of polar ice growth described exhibit not only changes in composition, but also in deposition rate. Figure~\ref{f:flux_trans} shows the growth rate as a function of the fraction of ice transported to the poles for several orbital configurations, with the end of the first migration stage indicated. As predicted from Figure~\ref{f:bar_plot}, obliquity is the major factor that determines what fraction of ice reaches the poles when a tropical reservoir is available (in Figure~\ref{f:flux_trans} the initial reservoir depletion occurs at higher ice fractions for higher obliquity). The North Polar flux experiences a decline of a factor $\sim$2 when reaching the second stage, i.e. when ice is no longer available in the tropics. The accumulation rate then continues to decrease as the source region of the ice moves polewards.  This means that if polar ice layers are derived from variations with constant periods, such as in Milankovitch cycles, layers sourced from higher latitudes will be thinner than those derived from tropical ice. For instance, at a low growth rate of 0.1 mm/Mars year, a layer of $\sim$2.5~m thickness forms in one 51 kyr precession cycle.

\begin{figure}[tb!]
    \centering
    \includegraphics[width=\linewidth]{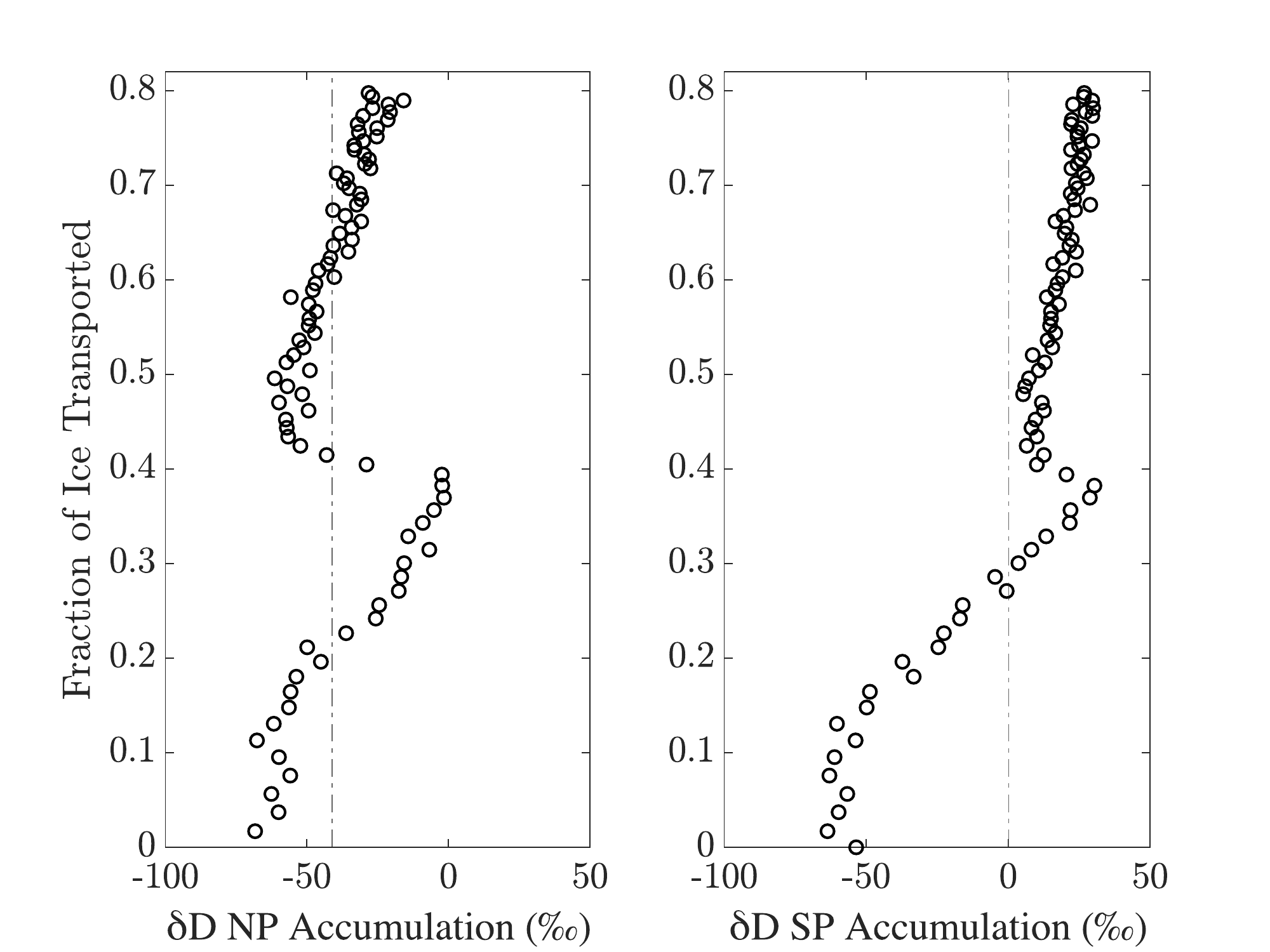}
    \caption{North and South Polar D/H profiles as a function of the fraction of ice transported for the same orbital configuration (obliquity of 25\deg{}, eccentricity of 0.093 and $L_p$ of 90\deg{}) and ice distribution as in Figure~\ref{f:surf_ice}. The data points are separated by one Mars year, and the dashed line is the mean D/H ratio of each final ice cap. Note that at later times (higher fraction of ice transported) the flux decreases (data points are more closely separated).  The variations in the profile result from the migration stages as well as the D/H ratio of the ice source. The south is enriched compared to the north, and the top-most layer of each cap is enriched compared to the bulk of that cap.}
    \label{f:Polar_profile_DH}
\end{figure}

With the accumulation rate and D/H ratio as functions of time, it is possible to construct a simulated ice stratigraphy. We combined these variables for the reference orbital configuration given in Figure~\ref{f:surf_ice}, to construct a polar profile  (Figure~\ref{f:Polar_profile_DH}). Different hydrogen isotopic sections appear along the profile due to the different migration stages and ice source locations mentioned above. The bottom portion of the profile (up to $\sim$0.2 ice fraction transported) is highly depleted in HDO relative to the equatorial source in both poles, due to the mid-high latitude reservoir fractionating the atmosphere as it grows. Once the tropical source is depleted, the enriched mid-high latitude deposit becomes the source, and the growing polar ice is therefore also enriched. As the polar ice grows at the expense of ever-higher latitude ice, its isotope ratio changes in accordance with the source, with a smaller modification due to heavy isotope trapping in intermediate latitudes between the source and the pole.
In both hemispheres, and for all profiles examined, our simulations show that the upper layers of each PLD is enriched in HDO compared to the average D/H ratio of that PLD (dashed line), though the enrichment is not always large. The effect is due to the HDO depletion of polar ice that accumulates early along with lower latitude enriched deposits. The simulations also show that the southern ice deposits are enriched in deuterium compared to the north. This north-south difference is robust across all simulations examined, which included different obliquities, eccentricities and perihelion alignments. An additional feature seen in the simulation is that the SPLD is consistently enriched in deuterium relative to the NPLD due to lower condensation temperatures caused by the topographic dichotomy. Note that the simulations capture the majority of PLD growth, but do not track any recent exchange that might occur after the ice has completely migrated to the polar regions above 75\deg{} latitude. If the top-most layer in the NPLD is sourced from southern deposits, it would be even more significantly enriched. The present-day atmosphere is primarily sourced from the northern ice cap, and we find its hydrogen isotopic composition is biased relative to the global ice reservoirs. Hence, the long-term water loss inferred from the current atmospheric deuterium concentration may deviate from the true value \citep{alsaeed2019,krasnopolsky2015,webster2013,villanueva2015}. 

\section{Summary and Discussion}
The goal of this work is to quantitatively describe the dependence of the physical and chemical stratigraphy of the polar layered deposits on orbital elements and surface ice distribution.
We ran simulations of the LMD-GCM with assumed past orbital elements configuration \citep{laskar2004} and an initial tropical ice reservoir placed in the equatorial region based on predictions of high obliquity simulations \citep{forget2006} and geologic evidence \citep{head2005}. 

When surface ice is available in the equatorial regions at low obliquity, the water vapor that sublimates from the tropics is in part transported to the polar region, with shorter and more intense polar summers enhancing the transport and increasing polar accumulation. The amount lost from the tropics can reach 10's mm PEL/Mars~year and is mostly determined by obliquity, 1-5 mm PEL/Mars year can reach each pole and the balance of the ice lost from the tropics is deposited in mid-high latitudes. Obliquity determines the sign of the NPLD flux, but the magnitude is also affected by eccentricity and $L_p$.  Hydrogen isotopic layering occurs when ice migrates from the equator to the poles, largely modulated by cycles of $L_p$ when a tropical reservoir is available. 

When comparing our results with those of \citet{levrard2007}, we find that increasing the model horizontal resolution and significantly extending the range of orbital states to also include different eccentricity and perihelion alignments reveal patterns of ice accumulation in the NPLD previously unrecognized. Specifically,
\citet{levrard2007} ran seven simulations with obliquities of 15\deg{} to 30\deg{} with restricted sets of orbital parameters and found that the NPLD accumulation rate is 1-2~mm PEL/Mars year. We show that accumulation at low to moderate obliquity is controlled by the intensity of summer (determined by eccentricity and perihelion position) and can reach 6~mm PEL/Mars year. We obtain similar results to \citet{levrard2007} for the NPLD loss rate at high obliquity, and for the threshold orbital configuration in which the NPLD transitions from accumulation to loss.

According to our simulations, the migration of ice from the tropics towards the poles occurs in two stages. A large fraction of the ice that ultimately reaches the polar caps is first deposited at mid-high latitudes. Once ice is depleted from the tropics, the mid-high latitude ice becomes the water vapor source and the flux to the poles decreases dramatically. In our simulations, the flux to the North Pole decreased to 0.1 mm/Mars year after $\sim$60\% of the ice that was placed in the tropics transported to the poles at obliquity of 15\deg. At this growth rate, a layer 2.5~m thick can form over a precession cycle. Since there are more 51~kyr precession cycles than 120~kyr obliquity cycles, this finding may help explain the large number of fine layers identified in the NPLD \citep{stozen2020,perron2009}. We expect even thinner layers after a higher fraction of ice has reached the poles at low obliquities, or if no more ice is available in the high latitudes and transport occurs only between the poles. An ice source below a dust lag layer could decrease the flux further \citep{bramson2019}. Past work identified two dominant layer thicknesses in the stratigraphic record $\sim$30 m and $\sim$1.6 m \citep{fishbaugh2010,perron2009,limaye2012,laskar2002,milkovich2005}.These thicknesses and their ratio are consistent with the two stages our simulations suggest. The thick layers are deposited when a tropical reservoir is available, and the thin layers when a tropical reservoir is absent.

Isotopic layering can shed light on past climatic conditions as well as aid in understanding the physical stratigraphy of the polar caps. Previous work \citep{vos2019} showed that an isotopic signal should be present in the cap based on a simple 3-box model that assumes fractionation on the surface with annual mean temperature. Here, we used GCM simulations with an HDO tracer that calculates the isotopic fractionation at each atmospheric grid cell and at each time step, allowing for geographic and seasonal variations \citep{rossi2021}. The results from the GCM support the previous conclusion and further show that migration stages affect the chemo-stratigraphy of the NPLD and SPLD. When a tropical reservoir is available, the hydrogen isotopic layering is only affected by the orbital elements, particularly $L_p$. At later times of migration, the D/H ratio of the ice that condenses in the polar regions is also affected by the surface ice distribution and the D/H ratio of the source ice. At each stage, the D/H enrichment in the mid-high latitudes is greatest at the most equatorial latitude at which ice condenses. Vapor condensing polewards of this margin is thus isotopically light. This result adds complexity to interpreting the expected chemo-stratigraphic record of the PLD. The bottom section (15-40\%  from the amount of ice present in the tropics according to Figure~\ref{f:flux_trans}) of both PLDs, which would have been deposited during the first stage when a tropical reservoir was present, is isotopically depleted, and subsequent deposits are enriched. 
The finding that the SPLD is isotopically enriched compared to the NPLD is robust for all orbital configurations in which ice is stable in the poles. Our results predict that the measured D/H composition of the present-day atmosphere, presumably dominated by contact with the NPLD top layers, is biased relative to the bulk ice in the polar caps, and the inferred fraction of water lost to space is likewise shifted.

Additional factors that were not considered in this study can affect the physio- and chemo- stratigraphy of the caps. These include the atmospheric dust distributions at different orbital configurations, regolith adsorption of water vapor, and time-dependent changes in the extent of subsurface ice over the timescales of orbital variations. These can alter the seasonal atmospheric water budget, the transport of water vapor, and the D/H fractionation factor. Cloud micro-physics and Rayleigh distillation in clouds may also affect the predicted evolution.


{\bf Corresponding author:}
Eran Vos, Department of Earth \& Planetary Sciences, Weizmann Institute of Science, Rehovot, 76100, Israel. 

{\bf Acknowledgments:}
We wish to thank the Helen Kimmel Center for Planetary Science and the Minerva Center for Life Under Extreme Planetary Conditions \#13599. The code for the LMD MGCM used for the simulations described in the text can be obtained in its entirety at http://svn.lmd.jussieu.fr/Planeto/trunk, and a detailed description of it can be found in the User Manual at https://www.lmd.jussieu.fr/$\sim$lmdz/planets/mars/user\_manual.pdf. The data made to create the figures is available online \citep{vos2022dataset}. No other data set was used in this work. The authors wish to thank Patricio Becerra and an anonymous reviewer for helpful comments, and Kevin Lewis and Jeremy Sotzen for valuable discussions.
\\

\bibliography{hdo}

\begin{thebibliography}{}

\bibitem [\protect \citeauthoryear {%
Alley%
}{%
Alley%
}{%
{\protect \APACyear {2000}}%
}]{%
alley2006}
\APACinsertmetastar {%
alley2006}%
\begin{APACrefauthors}%
Alley, R.%
\end{APACrefauthors}%
\ (\BED).
\unskip\
\newblock
\APACrefYear{2000}.
\newblock
\APACrefbtitle {The Two-Mile Time Machine: Ice Cores, Abrupt Climate Change,
  and Our Future} {The two-mile time machine: Ice cores, abrupt climate change,
  and our future}.
\newblock
\APACaddressPublisher{}{Princeton University Press}.
\newblock
\begin{APACrefDOI} \doi{10.2307/j.ctt6wq1f8} \end{APACrefDOI}
\PrintBackRefs{\CurrentBib}

\bibitem [\protect \citeauthoryear {%
Alsaeed%
\ \BBA {} Jakosky%
}{%
Alsaeed%
\ \BBA {} Jakosky%
}{%
{\protect \APACyear {2019}}%
}]{%
alsaeed2019}
\APACinsertmetastar {%
alsaeed2019}%
\begin{APACrefauthors}%
Alsaeed, N\BPBI R.%
\BCBT {}\ \BBA {} Jakosky, B\BPBI M.%
\end{APACrefauthors}%
\unskip\
\newblock
\APACrefYearMonthDay{2019}{}{}.
\newblock
{\BBOQ}\APACrefatitle {{M}ars Water and {D/H} Evolution From 3.3 {G}a to
  Present} {{M}ars water and {D/H} evolution from 3.3 {G}a to present}.{\BBCQ}
\newblock
\APACjournalVolNumPages{J.~Geophys. Res-Planet.}{124}{12}{3344--3353}.
\newblock
\begin{APACrefDOI} \doi{10.1029/2019JE006066} \end{APACrefDOI}
\PrintBackRefs{\CurrentBib}

\bibitem [\protect \citeauthoryear {%
Becerra%
, Sori%
\BCBL {}\ \BBA {} Byrne%
}{%
Becerra%
\ \protect \BOthers {.}}{%
{\protect \APACyear {2017}}%
}]{%
becerra2017}
\APACinsertmetastar {%
becerra2017}%
\begin{APACrefauthors}%
Becerra, P.%
, Sori, M\BPBI M.%
\BCBL {}\ \BBA {} Byrne, S.%
\end{APACrefauthors}%
\unskip\
\newblock
\APACrefYearMonthDay{2017}{}{}.
\newblock
{\BBOQ}\APACrefatitle {Signals of astronomical climate forcing in the exposure
  topography of the North Polar Layered Deposits of {M}ars} {Signals of
  astronomical climate forcing in the exposure topography of the north polar
  layered deposits of {M}ars}.{\BBCQ}
\newblock
\APACjournalVolNumPages{Geophys. Res. Lett.}{44}{1}{62-70}.
\newblock
\begin{APACrefDOI} \doi{https://doi.org/10.1002/2016GL071197} \end{APACrefDOI}
\PrintBackRefs{\CurrentBib}

\bibitem [\protect \citeauthoryear {%
Becerra%
\ \protect \BOthers {.}}{%
Becerra%
\ \protect \BOthers {.}}{%
{\protect \APACyear {2019}}%
}]{%
becerra2019}
\APACinsertmetastar {%
becerra2019}%
\begin{APACrefauthors}%
Becerra, P.%
, Sori, M\BPBI M.%
, Thomas, N.%
, Pommerol, A.%
, Simioni, E.%
, Sutton, S\BPBI S.%
\BDBL {}Cremonese, G.%
\end{APACrefauthors}%
\unskip\
\newblock
\APACrefYearMonthDay{2019}{}{}.
\newblock
{\BBOQ}\APACrefatitle {Timescales of the Climate Record in the South Polar Ice
  Cap of {M}ars} {Timescales of the climate record in the south polar ice cap
  of {M}ars}.{\BBCQ}
\newblock
\APACjournalVolNumPages{Geophys. Res. Lett.}{46}{13}{7268-7277}.
\newblock
\begin{APACrefDOI} \doi{https://doi.org/10.1029/2019GL083588} \end{APACrefDOI}
\PrintBackRefs{\CurrentBib}

\bibitem [\protect \citeauthoryear {%
Bramson%
, Byrne%
, Bapst%
, Smith%
\BCBL {}\ \BBA {} McClintock%
}{%
Bramson%
\ \protect \BOthers {.}}{%
{\protect \APACyear {2019}}%
}]{%
bramson2019}
\APACinsertmetastar {%
bramson2019}%
\begin{APACrefauthors}%
Bramson, A\BPBI M.%
, Byrne, S.%
, Bapst, J.%
, Smith, I\BPBI B.%
\BCBL {}\ \BBA {} McClintock, T.%
\end{APACrefauthors}%
\unskip\
\newblock
\APACrefYearMonthDay{2019}{}{}.
\newblock
{\BBOQ}\APACrefatitle {A Migration Model for the Polar Spiral Troughs of
  {M}ars} {A migration model for the polar spiral troughs of {M}ars}.{\BBCQ}
\newblock
\APACjournalVolNumPages{J.~Geophys. Res-Planet.}{124}{4}{1020-1043}.
\newblock
\begin{APACrefDOI} \doi{https://doi.org/10.1029/2018JE005806} \end{APACrefDOI}
\PrintBackRefs{\CurrentBib}

\bibitem [\protect \citeauthoryear {%
Byrne%
}{%
Byrne%
}{%
{\protect \APACyear {2009}}%
}]{%
byrne2009}
\APACinsertmetastar {%
byrne2009}%
\begin{APACrefauthors}%
Byrne, S.%
\end{APACrefauthors}%
\unskip\
\newblock
\APACrefYearMonthDay{2009}{}{}.
\newblock
{\BBOQ}\APACrefatitle {The Polar Deposits of {M}ars} {The polar deposits of
  {M}ars}.{\BBCQ}
\newblock
\APACjournalVolNumPages{Annu. Rev. Earth. Pl. Sc.}{37}{1}{535-560}.
\newblock
\begin{APACrefDOI} \doi{0.1146/annurev.earth.031208.100101} \end{APACrefDOI}
\PrintBackRefs{\CurrentBib}

\bibitem [\protect \citeauthoryear {%
Cutts%
\ \BBA {} Lewis%
}{%
Cutts%
\ \BBA {} Lewis%
}{%
{\protect \APACyear {1982}}%
}]{%
cutts1982}
\APACinsertmetastar {%
cutts1982}%
\begin{APACrefauthors}%
Cutts, J\BPBI A.%
\BCBT {}\ \BBA {} Lewis, B\BPBI H.%
\end{APACrefauthors}%
\unskip\
\newblock
\APACrefYearMonthDay{1982}{}{}.
\newblock
{\BBOQ}\APACrefatitle {Models of climate cycles recorded in Martian polar
  layered deposits} {Models of climate cycles recorded in martian polar layered
  deposits}.{\BBCQ}
\newblock
\APACjournalVolNumPages{Icarus}{50}{2}{216-244}.
\newblock
\begin{APACrefDOI} \doi{https://doi.org/10.1016/0019-1035(82)90124-5}
  \end{APACrefDOI}
\PrintBackRefs{\CurrentBib}

\bibitem [\protect \citeauthoryear {%
Emmett%
, Murphy%
\BCBL {}\ \BBA {} Kahre%
}{%
Emmett%
\ \protect \BOthers {.}}{%
{\protect \APACyear {2020}}%
}]{%
emmett2020}
\APACinsertmetastar {%
emmett2020}%
\begin{APACrefauthors}%
Emmett, J.%
, Murphy, J.%
\BCBL {}\ \BBA {} Kahre, M.%
\end{APACrefauthors}%
\unskip\
\newblock
\APACrefYearMonthDay{2020}{}{}.
\newblock
{\BBOQ}\APACrefatitle {Obliquity dependence of the formation of the martian
  polar layered deposits} {Obliquity dependence of the formation of the martian
  polar layered deposits}.{\BBCQ}
\newblock
\APACjournalVolNumPages{Planet. Space. Sci}{193}{105047}{}.
\newblock
\begin{APACrefDOI} \doi{10.1016/j.pss.2020.105047} \end{APACrefDOI}
\PrintBackRefs{\CurrentBib}

\bibitem [\protect \citeauthoryear {%
Fenton%
\ \BBA {} Herkenhoff%
}{%
Fenton%
\ \BBA {} Herkenhoff%
}{%
{\protect \APACyear {2000}}%
}]{%
fenton2000}
\APACinsertmetastar {%
fenton2000}%
\begin{APACrefauthors}%
Fenton, L.%
\BCBT {}\ \BBA {} Herkenhoff, K.%
\end{APACrefauthors}%
\unskip\
\newblock
\APACrefYearMonthDay{2000}{}{}.
\newblock
{\BBOQ}\APACrefatitle {Topography and Stratigraphy of the Northern Martian
  Polar Layered Deposits Using Photoclinometry, Stereogrammetry, and MOLA
  Altimetry} {Topography and stratigraphy of the northern martian polar layered
  deposits using photoclinometry, stereogrammetry, and mola altimetry}.{\BBCQ}
\newblock
\APACjournalVolNumPages{Icarus}{147}{2}{433--443}.
\newblock
\begin{APACrefDOI} \doi{10.1006/icar.2000.6459} \end{APACrefDOI}
\PrintBackRefs{\CurrentBib}

\bibitem [\protect \citeauthoryear {%
Fishbaugh%
\ \BBA {} Hvidberg%
}{%
Fishbaugh%
\ \BBA {} Hvidberg%
}{%
{\protect \APACyear {2006}}%
}]{%
fishbaugh2006}
\APACinsertmetastar {%
fishbaugh2006}%
\begin{APACrefauthors}%
Fishbaugh, K\BPBI E.%
\BCBT {}\ \BBA {} Hvidberg, C\BPBI S.%
\end{APACrefauthors}%
\unskip\
\newblock
\APACrefYearMonthDay{2006}{}{}.
\newblock
{\BBOQ}\APACrefatitle {Martian north polar layered deposits stratigraphy:
  Implications for accumulation rates and flow} {Martian north polar layered
  deposits stratigraphy: Implications for accumulation rates and flow}.{\BBCQ}
\newblock
\APACjournalVolNumPages{J.~Geophys. Res-Planet.}{111}{E6}{}.
\newblock
\begin{APACrefDOI} \doi{10.1029/2005JE002571} \end{APACrefDOI}
\PrintBackRefs{\CurrentBib}

\bibitem [\protect \citeauthoryear {%
Fishbaugh%
\ \protect \BOthers {.}}{%
Fishbaugh%
\ \protect \BOthers {.}}{%
{\protect \APACyear {2010}}%
}]{%
fishbaugh2010}
\APACinsertmetastar {%
fishbaugh2010}%
\begin{APACrefauthors}%
Fishbaugh, K\BPBI E.%
, Hvidberg, C\BPBI S.%
, Byrne, S.%
, Russell, P\BPBI S.%
, Herkenhoff, K\BPBI E.%
, Winstrup, M.%
\BCBL {}\ \BBA {} Kirk, R.%
\end{APACrefauthors}%
\unskip\
\newblock
\APACrefYearMonthDay{2010}{}{}.
\newblock
{\BBOQ}\APACrefatitle {First high-resolution stratigraphic column of the
  Martian north polar layered deposits} {First high-resolution stratigraphic
  column of the martian north polar layered deposits}.{\BBCQ}
\newblock
\APACjournalVolNumPages{Geophys. Res. Lett.}{37}{7}{}.
\newblock
\begin{APACrefDOI} \doi{https://doi.org/10.1029/2009GL041642} \end{APACrefDOI}
\PrintBackRefs{\CurrentBib}

\bibitem [\protect \citeauthoryear {%
Forget%
, Haberle%
, F.Montmessin%
, Levrard%
\BCBL {}\ \BBA {} Head%
}{%
Forget%
\ \protect \BOthers {.}}{%
{\protect \APACyear {2006}}%
}]{%
forget2006}
\APACinsertmetastar {%
forget2006}%
\begin{APACrefauthors}%
Forget, F.%
, Haberle, R\BPBI M.%
, F.Montmessin%
, Levrard, B.%
\BCBL {}\ \BBA {} Head, J\BPBI W.%
\end{APACrefauthors}%
\unskip\
\newblock
\APACrefYearMonthDay{2006}{}{}.
\newblock
{\BBOQ}\APACrefatitle {Formation of Glaciers on {M}ars by Atmospheric
  Precipitation at High Obliquity} {Formation of glaciers on {M}ars by
  atmospheric precipitation at high obliquity}.{\BBCQ}
\newblock
\APACjournalVolNumPages{Science}{311}{}{368--371}.
\newblock
\begin{APACrefDOI} \doi{10.1126/science.1120335} \end{APACrefDOI}
\PrintBackRefs{\CurrentBib}

\bibitem [\protect \citeauthoryear {%
Forget%
\ \protect \BOthers {.}}{%
Forget%
\ \protect \BOthers {.}}{%
{\protect \APACyear {1999}}%
}]{%
forget1999}
\APACinsertmetastar {%
forget1999}%
\begin{APACrefauthors}%
Forget, F.%
, Hourdin, F.%
, Fournier, R.%
, Hourdin, C.%
, Talagrand, O.%
, Collins, M.%
\BDBL {}Huot, J\BHBI P.%
\end{APACrefauthors}%
\unskip\
\newblock
\APACrefYearMonthDay{1999}{}{}.
\newblock
{\BBOQ}\APACrefatitle {Improved general circulation models of the Martian
  atmosphere from the surface to above 80 km} {Improved general circulation
  models of the martian atmosphere from the surface to above 80 km}.{\BBCQ}
\newblock
\APACjournalVolNumPages{J.~Geophys. Res-Planet.}{104}{E10}{}.
\newblock
\begin{APACrefDOI} \doi{10.1029/1999JE001025} \end{APACrefDOI}
\PrintBackRefs{\CurrentBib}

\bibitem [\protect \citeauthoryear {%
Head%
\ \protect \BOthers {.}}{%
Head%
\ \protect \BOthers {.}}{%
{\protect \APACyear {2005}}%
}]{%
head2005}
\APACinsertmetastar {%
head2005}%
\begin{APACrefauthors}%
Head, J\BPBI W.%
, Neukum, G.%
, aumann, R.%
, Hiesinger, H.%
, Hauber, E.%
, Carr, M.%
\BDBL {}Team, T\BPBI H\BPBI C\BHBI I.%
\end{APACrefauthors}%
\unskip\
\newblock
\APACrefYearMonthDay{2005}{}{}.
\newblock
{\BBOQ}\APACrefatitle {Tropical to mid-latitude snow and ice accumulation, flow
  and glaciation on {M}ars} {Tropical to mid-latitude snow and ice
  accumulation, flow and glaciation on {M}ars}.{\BBCQ}
\newblock
\APACjournalVolNumPages{Nature}{434}{}{346--351}.
\newblock
\begin{APACrefDOI} \doi{10.1038/nature03359} \end{APACrefDOI}
\PrintBackRefs{\CurrentBib}

\bibitem [\protect \citeauthoryear {%
Hvidberg%
\ \protect \BOthers {.}}{%
Hvidberg%
\ \protect \BOthers {.}}{%
{\protect \APACyear {2012}}%
}]{%
hvidberg2012}
\APACinsertmetastar {%
hvidberg2012}%
\begin{APACrefauthors}%
Hvidberg, C.%
, Fishbaugh, K.%
, Winstrup, M.%
, Svensson, A.%
, Byrne, S.%
\BCBL {}\ \BBA {} Herkenhoff, K.%
\end{APACrefauthors}%
\unskip\
\newblock
\APACrefYearMonthDay{2012}{}{}.
\newblock
{\BBOQ}\APACrefatitle {Reading the climate record of the martian polar layered
  deposits} {Reading the climate record of the martian polar layered
  deposits}.{\BBCQ}
\newblock
\APACjournalVolNumPages{Icarus}{221}{1}{405--419}.
\newblock
\begin{APACrefDOI} \doi{10.1016/j.icarus.2012.08.009} \end{APACrefDOI}
\PrintBackRefs{\CurrentBib}

\bibitem [\protect \citeauthoryear {%
Jakosky%
\ \protect \BOthers {.}}{%
Jakosky%
\ \protect \BOthers {.}}{%
{\protect \APACyear {2018}}%
}]{%
jakosky2018}
\APACinsertmetastar {%
jakosky2018}%
\begin{APACrefauthors}%
Jakosky, B.%
, Brain, D.%
, Chaffin, M.%
, Curry, S.%
, Deighan, J.%
, Grebowsky, J.%
\BDBL {}Zurek, R.%
\end{APACrefauthors}%
\unskip\
\newblock
\APACrefYearMonthDay{2018}{}{}.
\newblock
{\BBOQ}\APACrefatitle {Loss of the Martian atmosphere to space: Present-day
  loss rates determined from MAVEN observations and integrated loss through
  time} {Loss of the martian atmosphere to space: Present-day loss rates
  determined from maven observations and integrated loss through time}.{\BBCQ}
\newblock
\APACjournalVolNumPages{Icarus}{315}{}{146--157}.
\newblock
\begin{APACrefDOI} \doi{10.1016/j.icarus.2018.05.030} \end{APACrefDOI}
\PrintBackRefs{\CurrentBib}

\bibitem [\protect \citeauthoryear {%
Krasnopolsky%
}{%
Krasnopolsky%
}{%
{\protect \APACyear {2015}}%
}]{%
krasnopolsky2015}
\APACinsertmetastar {%
krasnopolsky2015}%
\begin{APACrefauthors}%
Krasnopolsky, V.%
\end{APACrefauthors}%
\unskip\
\newblock
\APACrefYearMonthDay{2015}{}{}.
\newblock
{\BBOQ}\APACrefatitle {Variations of the HDO/H2O ratio in the martian
  atmosphere and loss of water from {M}ars} {Variations of the hdo/h2o ratio in
  the martian atmosphere and loss of water from {M}ars}.{\BBCQ}
\newblock
\APACjournalVolNumPages{Icarus}{257}{}{377--386}.
\newblock
\begin{APACrefDOI} \doi{10.1016/j.icarus.2015.05.021.} \end{APACrefDOI}
\PrintBackRefs{\CurrentBib}

\bibitem [\protect \citeauthoryear {%
Lamb%
\ \protect \BOthers {.}}{%
Lamb%
\ \protect \BOthers {.}}{%
{\protect \APACyear {2017}}%
}]{%
lamb2017}
\APACinsertmetastar {%
lamb2017}%
\begin{APACrefauthors}%
Lamb, K\BPBI D.%
, Clouser, B\BPBI W.%
, Bolot, M.%
, Sarkozy, L.%
, Ebert, V.%
, Saathoff, H.%
\BDBL {}Moyer, E\BPBI J.%
\end{APACrefauthors}%
\unskip\
\newblock
\APACrefYearMonthDay{2017}{}{}.
\newblock
{\BBOQ}\APACrefatitle {Laboratory measurements of HDO/H2O isotopic
  fractionation during ice deposition in simulated cirrus clouds} {Laboratory
  measurements of hdo/h2o isotopic fractionation during ice deposition in
  simulated cirrus clouds}.{\BBCQ}
\newblock
\APACjournalVolNumPages{P. Natl. Acad. Sci.}{114}{22}{5612--5617}.
\newblock
\begin{APACrefDOI} \doi{10.1073/pnas.1618374114} \end{APACrefDOI}
\PrintBackRefs{\CurrentBib}

\bibitem [\protect \citeauthoryear {%
Laskar%
, Levrard%
\BCBL {}\ \BBA {} Mustard%
}{%
Laskar%
\ \protect \BOthers {.}}{%
{\protect \APACyear {2002}}%
}]{%
laskar2002}
\APACinsertmetastar {%
laskar2002}%
\begin{APACrefauthors}%
Laskar, J.%
, Levrard, B.%
\BCBL {}\ \BBA {} Mustard, J.%
\end{APACrefauthors}%
\unskip\
\newblock
\APACrefYearMonthDay{2002}{}{}.
\newblock
{\BBOQ}\APACrefatitle {Orbital forcing of the martian polar layered deposits}
  {Orbital forcing of the martian polar layered deposits}.{\BBCQ}
\newblock
\APACjournalVolNumPages{Nature}{419}{}{375--377}.
\newblock
\begin{APACrefDOI} \doi{10.1038/nature01066} \end{APACrefDOI}
\PrintBackRefs{\CurrentBib}

\bibitem [\protect \citeauthoryear {%
Laskar%
\ \protect \BOthers {.}}{%
Laskar%
\ \protect \BOthers {.}}{%
{\protect \APACyear {2004}}%
}]{%
laskar2004}
\APACinsertmetastar {%
laskar2004}%
\begin{APACrefauthors}%
Laskar, J.%
, Robutel, P.%
, Joutel, F.%
, Gastineau, M.%
, Correia, A\BPBI C\BPBI M.%
\BCBL {}\ \BBA {} Levrard, B.%
\end{APACrefauthors}%
\unskip\
\newblock
\APACrefYearMonthDay{2004}{}{}.
\newblock
{\BBOQ}\APACrefatitle {A long-term numerical solution for the insolation
  quantities of the Earth} {A long-term numerical solution for the insolation
  quantities of the earth}.{\BBCQ}
\newblock
\APACjournalVolNumPages{Astron. Astrophys.}{428}{1}{261--285}.
\newblock
\begin{APACrefDOI} \doi{10.1051/0004-6361:20041335} \end{APACrefDOI}
\PrintBackRefs{\CurrentBib}

\bibitem [\protect \citeauthoryear {%
Levrard%
, Forget%
, Montmessin%
\BCBL {}\ \BBA {} Laskar%
}{%
Levrard%
\ \protect \BOthers {.}}{%
{\protect \APACyear {2004}}%
}]{%
levrard2004}
\APACinsertmetastar {%
levrard2004}%
\begin{APACrefauthors}%
Levrard, B.%
, Forget, F.%
, Montmessin, F.%
\BCBL {}\ \BBA {} Laskar, J.%
\end{APACrefauthors}%
\unskip\
\newblock
\APACrefYearMonthDay{2004}{}{}.
\newblock
{\BBOQ}\APACrefatitle {Recent ice-rich deposits formed at high latitudes on
  {M}ars by sublimation of unstable equatorial ice during low obliquity}
  {Recent ice-rich deposits formed at high latitudes on {M}ars by sublimation
  of unstable equatorial ice during low obliquity}.{\BBCQ}
\newblock
\APACjournalVolNumPages{Nature}{431}{}{1072--1075}.
\newblock
\begin{APACrefDOI} \doi{10.1038/nature03055} \end{APACrefDOI}
\PrintBackRefs{\CurrentBib}

\bibitem [\protect \citeauthoryear {%
Levrard%
, Forget%
, Montmessin%
\BCBL {}\ \BBA {} Laskar%
}{%
Levrard%
\ \protect \BOthers {.}}{%
{\protect \APACyear {2007}}%
}]{%
levrard2007}
\APACinsertmetastar {%
levrard2007}%
\begin{APACrefauthors}%
Levrard, B.%
, Forget, F.%
, Montmessin, F.%
\BCBL {}\ \BBA {} Laskar, J.%
\end{APACrefauthors}%
\unskip\
\newblock
\APACrefYearMonthDay{2007}{}{}.
\newblock
{\BBOQ}\APACrefatitle {Recent formation and evolution of northern Martian polar
  layered deposits as inferred from a Global Climate Model} {Recent formation
  and evolution of northern martian polar layered deposits as inferred from a
  global climate model}.{\BBCQ}
\newblock
\APACjournalVolNumPages{J.~Geophys. Res-Planet.}{112}{E6}{}.
\newblock
\begin{APACrefDOI} \doi{10.1029/2006JE002772} \end{APACrefDOI}
\PrintBackRefs{\CurrentBib}

\bibitem [\protect \citeauthoryear {%
Limaye%
, Aharonson%
\BCBL {}\ \BBA {} Perron%
}{%
Limaye%
\ \protect \BOthers {.}}{%
{\protect \APACyear {2012}}%
}]{%
limaye2012}
\APACinsertmetastar {%
limaye2012}%
\begin{APACrefauthors}%
Limaye, A\BPBI B\BPBI S.%
, Aharonson, O.%
\BCBL {}\ \BBA {} Perron, J\BPBI T.%
\end{APACrefauthors}%
\unskip\
\newblock
\APACrefYearMonthDay{2012}{}{}.
\newblock
{\BBOQ}\APACrefatitle {Detailed stratigraphy and bed thickness of the {M}ars
  north and south polar layered deposits} {Detailed stratigraphy and bed
  thickness of the {M}ars north and south polar layered deposits}.{\BBCQ}
\newblock
\APACjournalVolNumPages{J.~Geophys. Res-Planet.}{117}{E6}{}.
\newblock
\begin{APACrefDOI} \doi{https://doi.org/10.1029/2011JE003961} \end{APACrefDOI}
\PrintBackRefs{\CurrentBib}

\bibitem [\protect \citeauthoryear {%
Madeleine%
\ \protect \BOthers {.}}{%
Madeleine%
\ \protect \BOthers {.}}{%
{\protect \APACyear {2009}}%
}]{%
madeleine2009}
\APACinsertmetastar {%
madeleine2009}%
\begin{APACrefauthors}%
Madeleine, J\BHBI B.%
, Forget, F.%
, Head, J\BPBI W.%
, Levrard, B.%
, Montmessin, F.%
\BCBL {}\ \BBA {} Millour, E.%
\end{APACrefauthors}%
\unskip\
\newblock
\APACrefYearMonthDay{2009}{}{}.
\newblock
{\BBOQ}\APACrefatitle {Amazonian northern mid-latitude glaciation on {M}ars: A
  proposed climate scenario} {Amazonian northern mid-latitude glaciation on
  {M}ars: A proposed climate scenario}.{\BBCQ}
\newblock
\APACjournalVolNumPages{Icarus}{203}{2}{390--405}.
\newblock
\begin{APACrefDOI} \doi{10.1016/j.icarus.2009.04.037} \end{APACrefDOI}
\PrintBackRefs{\CurrentBib}

\bibitem [\protect \citeauthoryear {%
Madeleine%
, Forget%
, Millour%
, Montabone%
\BCBL {}\ \BBA {} Wolff%
}{%
Madeleine%
\ \protect \BOthers {.}}{%
{\protect \APACyear {2011}}%
}]{%
madeleine2011}
\APACinsertmetastar {%
madeleine2011}%
\begin{APACrefauthors}%
Madeleine, J\BHBI B.%
, Forget, F.%
, Millour, E.%
, Montabone, L.%
\BCBL {}\ \BBA {} Wolff, M\BPBI J.%
\end{APACrefauthors}%
\unskip\
\newblock
\APACrefYearMonthDay{2011}{}{}.
\newblock
{\BBOQ}\APACrefatitle {Revisiting the radiative impact of dust on {M}ars using
  the LMD Global Climate Model} {Revisiting the radiative impact of dust on
  {M}ars using the lmd global climate model}.{\BBCQ}
\newblock
\APACjournalVolNumPages{J.~Geophys. Res-Planet.}{116}{E11}{}.
\newblock
\begin{APACrefDOI} \doi{https://doi.org/10.1029/2011JE003855} \end{APACrefDOI}
\PrintBackRefs{\CurrentBib}

\bibitem [\protect \citeauthoryear {%
Madeleine%
, Forget%
, Millour%
, Navarro%
\BCBL {}\ \BBA {} Spiga%
}{%
Madeleine%
\ \protect \BOthers {.}}{%
{\protect \APACyear {2012}}%
}]{%
madeleine2012}
\APACinsertmetastar {%
madeleine2012}%
\begin{APACrefauthors}%
Madeleine, J\BHBI B.%
, Forget, F.%
, Millour, E.%
, Navarro, T.%
\BCBL {}\ \BBA {} Spiga, A.%
\end{APACrefauthors}%
\unskip\
\newblock
\APACrefYearMonthDay{2012}{}{}.
\newblock
{\BBOQ}\APACrefatitle {The influence of radiatively active water ice clouds on
  the Martian climate} {The influence of radiatively active water ice clouds on
  the martian climate}.{\BBCQ}
\newblock
\APACjournalVolNumPages{Geophys. Res. Lett.}{39}{23}{}.
\newblock
\begin{APACrefDOI} \doi{https://doi.org/10.1029/2012GL053564} \end{APACrefDOI}
\PrintBackRefs{\CurrentBib}

\bibitem [\protect \citeauthoryear {%
Madeleine%
\ \protect \BOthers {.}}{%
Madeleine%
\ \protect \BOthers {.}}{%
{\protect \APACyear {2014}}%
}]{%
madeleine2014}
\APACinsertmetastar {%
madeleine2014}%
\begin{APACrefauthors}%
Madeleine, J\BHBI B.%
, Head, J\BPBI W.%
, Forget, F.%
, Navarro, T.%
, Millour, E.%
, Spiga, A.%
\BDBL {}Dickson, J\BPBI L.%
\end{APACrefauthors}%
\unskip\
\newblock
\APACrefYearMonthDay{2014}{}{}.
\newblock
{\BBOQ}\APACrefatitle {Recent Ice Ages on {M}ars: The role of radiatively
  active clouds and cloud microphysics} {Recent ice ages on {M}ars: The role of
  radiatively active clouds and cloud microphysics}.{\BBCQ}
\newblock
\APACjournalVolNumPages{Geophys. Res. Lett.}{41}{14}{4873-4879}.
\newblock
\begin{APACrefDOI} \doi{https://doi.org/10.1002/2014GL059861} \end{APACrefDOI}
\PrintBackRefs{\CurrentBib}

\bibitem [\protect \citeauthoryear {%
Merlivat%
\ \BBA {} Nief%
}{%
Merlivat%
\ \BBA {} Nief%
}{%
{\protect \APACyear {1967}}%
}]{%
merlivat1967}
\APACinsertmetastar {%
merlivat1967}%
\begin{APACrefauthors}%
Merlivat, L.%
\BCBT {}\ \BBA {} Nief, G.%
\end{APACrefauthors}%
\unskip\
\newblock
\APACrefYearMonthDay{1967}{}{}.
\newblock
{\BBOQ}\APACrefatitle {Fractionnement isotopique lors des changements d‘état
  solide-vapeur et liquide-vapeur de l'eau à des températures inférieures à
  0$^{\circ}$ C} {Fractionnement isotopique lors des changements d‘état
  solide-vapeur et liquide-vapeur de l'eau à des températures inférieures à
  0$^{\circ}$ c}.{\BBCQ}
\newblock
\APACjournalVolNumPages{Tellus}{19}{1}{}.
\newblock
\begin{APACrefDOI} \doi{10.1111/j.2153-3490.1967.tb01465.x} \end{APACrefDOI}
\PrintBackRefs{\CurrentBib}

\bibitem [\protect \citeauthoryear {%
Milkovich%
\ \BBA {} Head%
}{%
Milkovich%
\ \BBA {} Head%
}{%
{\protect \APACyear {2005}}%
}]{%
milkovich2005}
\APACinsertmetastar {%
milkovich2005}%
\begin{APACrefauthors}%
Milkovich, S\BPBI M.%
\BCBT {}\ \BBA {} Head, J.%
\end{APACrefauthors}%
\unskip\
\newblock
\APACrefYearMonthDay{2005}{}{}.
\newblock
{\BBOQ}\APACrefatitle {North polar cap of {M}ars: Polar layered deposit
  characterization and identification of a fundamental climate signal} {North
  polar cap of {M}ars: Polar layered deposit characterization and
  identification of a fundamental climate signal}.{\BBCQ}
\newblock
\APACjournalVolNumPages{J.~Geophys. Res-Planet.}{110}{E1}{}.
\newblock
\begin{APACrefDOI} \doi{10.1029/2004JE002349} \end{APACrefDOI}
\PrintBackRefs{\CurrentBib}

\bibitem [\protect \citeauthoryear {%
Milkovich%
\ \BBA {} Plaut%
}{%
Milkovich%
\ \BBA {} Plaut%
}{%
{\protect \APACyear {2008}}%
}]{%
milkovich2008}
\APACinsertmetastar {%
milkovich2008}%
\begin{APACrefauthors}%
Milkovich, S\BPBI M.%
\BCBT {}\ \BBA {} Plaut, J\BPBI J.%
\end{APACrefauthors}%
\unskip\
\newblock
\APACrefYearMonthDay{2008}{}{}.
\newblock
{\BBOQ}\APACrefatitle {Martian South Polar Layered Deposit stratigraphy and
  implications for accumulation history} {Martian south polar layered deposit
  stratigraphy and implications for accumulation history}.{\BBCQ}
\newblock
\APACjournalVolNumPages{J.~Geophys. Res-Planet.}{113}{E6}{}.
\newblock
\begin{APACrefDOI} \doi{https://doi.org/10.1029/2007JE002987} \end{APACrefDOI}
\PrintBackRefs{\CurrentBib}

\bibitem [\protect \citeauthoryear {%
{Millour}%
\ \protect \BOthers {.}}{%
{Millour}%
\ \protect \BOthers {.}}{%
{\protect \APACyear {2018}}%
}]{%
millour2018}
\APACinsertmetastar {%
millour2018}%
\begin{APACrefauthors}%
{Millour}, E.%
, {Forget}, F.%
, {Spiga}, A.%
, {Vals}, M.%
, {Zakharov}, V.%
, {Montabone}, L.%
\BDBL {}{MCD Development Team}%
\end{APACrefauthors}%
\unskip\
\newblock
\APACrefYearMonthDay{2018}{{\APACmonth{02}}}{}.
\newblock
{\BBOQ}\APACrefatitle {{The {M}ars Climate Database (version 5.3)}} {{The
  {M}ars Climate Database (version 5.3)}}.{\BBCQ}
\newblock
\BIn{} \APACrefbtitle {From Mars Express to ExoMars} {From mars express to
  exomars}\ (\BPG~68).
\PrintBackRefs{\CurrentBib}

\bibitem [\protect \citeauthoryear {%
Montmessin%
, Forget%
, Rannou%
, Cabane%
\BCBL {}\ \BBA {} Haberle%
}{%
Montmessin%
\ \protect \BOthers {.}}{%
{\protect \APACyear {2004}}%
}]{%
montmessin2004}
\APACinsertmetastar {%
montmessin2004}%
\begin{APACrefauthors}%
Montmessin, F.%
, Forget, F.%
, Rannou, P.%
, Cabane, M.%
\BCBL {}\ \BBA {} Haberle, R\BPBI M.%
\end{APACrefauthors}%
\unskip\
\newblock
\APACrefYearMonthDay{2004}{}{}.
\newblock
{\BBOQ}\APACrefatitle {Origin and role of water ice clouds in the Martian water
  cycle as inferred from a general circulation model} {Origin and role of water
  ice clouds in the martian water cycle as inferred from a general circulation
  model}.{\BBCQ}
\newblock
\APACjournalVolNumPages{J.~Geophys. Res-Planet.}{109}{E10}{}.
\newblock
\begin{APACrefDOI} \doi{10.1029/2004JE002284} \end{APACrefDOI}
\PrintBackRefs{\CurrentBib}

\bibitem [\protect \citeauthoryear {%
Montmessin%
, Fouchet%
\BCBL {}\ \BBA {} Forget%
}{%
Montmessin%
\ \protect \BOthers {.}}{%
{\protect \APACyear {2005}}%
}]{%
montmessin2005}
\APACinsertmetastar {%
montmessin2005}%
\begin{APACrefauthors}%
Montmessin, F.%
, Fouchet, T.%
\BCBL {}\ \BBA {} Forget, F.%
\end{APACrefauthors}%
\unskip\
\newblock
\APACrefYearMonthDay{2005}{}{}.
\newblock
{\BBOQ}\APACrefatitle {Modeling the annual cycle of HDO in the Martian
  atmosphere} {Modeling the annual cycle of hdo in the martian
  atmosphere}.{\BBCQ}
\newblock
\APACjournalVolNumPages{J.~Geophys. Res-Planet.}{110}{E3}{}.
\newblock
\begin{APACrefDOI} \doi{10.1029/2004JE002357} \end{APACrefDOI}
\PrintBackRefs{\CurrentBib}

\bibitem [\protect \citeauthoryear {%
Perron%
\ \BBA {} Huybers%
}{%
Perron%
\ \BBA {} Huybers%
}{%
{\protect \APACyear {2009}}%
}]{%
perron2009}
\APACinsertmetastar {%
perron2009}%
\begin{APACrefauthors}%
Perron, T.%
\BCBT {}\ \BBA {} Huybers, P.%
\end{APACrefauthors}%
\unskip\
\newblock
\APACrefYearMonthDay{2009}{}{}.
\newblock
{\BBOQ}\APACrefatitle {Is there an orbital signal in the polar layered deposits
  on {M}ars?} {Is there an orbital signal in the polar layered deposits on
  {M}ars?}{\BBCQ}
\newblock
\APACjournalVolNumPages{Geology}{37}{2}{155--158}.
\newblock
\begin{APACrefDOI} \doi{10.1130/G25143A.1} \end{APACrefDOI}
\PrintBackRefs{\CurrentBib}

\bibitem [\protect \citeauthoryear {%
Plaut%
\ \protect \BOthers {.}}{%
Plaut%
\ \protect \BOthers {.}}{%
{\protect \APACyear {2007}}%
}]{%
plaut2007}
\APACinsertmetastar {%
plaut2007}%
\begin{APACrefauthors}%
Plaut, J\BPBI J.%
, Picardi, G.%
, Safaeinili, A.%
, Ivanov, A\BPBI B.%
, Milkovich, S\BPBI M.%
, Cicchetti, A.%
\BDBL {}Edenhofer, P.%
\end{APACrefauthors}%
\unskip\
\newblock
\APACrefYearMonthDay{2007}{}{}.
\newblock
{\BBOQ}\APACrefatitle {Subsurface Radar Sounding of the South Polar Layered
  Deposits of {M}ars} {Subsurface radar sounding of the south polar layered
  deposits of {M}ars}.{\BBCQ}
\newblock
\APACjournalVolNumPages{Science}{316}{5821}{92--95}.
\newblock
\begin{APACrefDOI} \doi{10.1126/science.1139672} \end{APACrefDOI}
\PrintBackRefs{\CurrentBib}

\bibitem [\protect \citeauthoryear {%
Richardson%
\ \BBA {} Wilson%
}{%
Richardson%
\ \BBA {} Wilson%
}{%
{\protect \APACyear {2002}}%
}]{%
Richardson2002}
\APACinsertmetastar {%
Richardson2002}%
\begin{APACrefauthors}%
Richardson, M.%
\BCBT {}\ \BBA {} Wilson, J.%
\end{APACrefauthors}%
\unskip\
\newblock
\APACrefYearMonthDay{2002}{}{}.
\newblock
{\BBOQ}\APACrefatitle {A topographically forced asymmetry in the martian
  circulation and climate} {A topographically forced asymmetry in the martian
  circulation and climate}.{\BBCQ}
\newblock
\APACjournalVolNumPages{Nature}{416}{}{298--301}.
\newblock
\begin{APACrefDOI} \doi{10.1038/416298a} \end{APACrefDOI}
\PrintBackRefs{\CurrentBib}

\bibitem [\protect \citeauthoryear {%
Rossi%
\ \protect \BOthers {.}}{%
Rossi%
\ \protect \BOthers {.}}{%
{\protect \APACyear {2021}}%
}]{%
rossi2021}
\APACinsertmetastar {%
rossi2021}%
\begin{APACrefauthors}%
Rossi, L.%
, Vals, M.%
, Montmessin, F.%
, Forget, F.%
, Millour, E.%
, Fedorova, A.%
\BDBL {}Korablev, O.%
\end{APACrefauthors}%
\unskip\
\newblock
\APACrefYearMonthDay{2021}{}{}.
\newblock
{\BBOQ}\APACrefatitle {The effect of the Martian 2018 global dust storm on
  {HDO} as predicted by a {M}ars {G}lobal {C}limate {M}odel} {The effect of the
  martian 2018 global dust storm on {HDO} as predicted by a {M}ars {G}lobal
  {C}limate {M}odel}.{\BBCQ}
\newblock
\APACjournalVolNumPages{Geophys. Res. Lett.}{}{}{}.
\newblock
\begin{APACrefDOI} \doi{10.1029/2020GL090962} \end{APACrefDOI}
\PrintBackRefs{\CurrentBib}

\bibitem [\protect \citeauthoryear {%
Smith%
\ \protect \BOthers {.}}{%
Smith%
\ \protect \BOthers {.}}{%
{\protect \APACyear {2020}}%
}]{%
smith2020}
\APACinsertmetastar {%
smith2020}%
\begin{APACrefauthors}%
Smith, I\BPBI B.%
, Hayne, P\BPBI O.%
, Byrne, S.%
, Becerra, P.%
, Kahre, M.%
, Calvin, W.%
\BDBL {}Siegler, M.%
\end{APACrefauthors}%
\unskip\
\newblock
\APACrefYearMonthDay{2020}{}{}.
\newblock
{\BBOQ}\APACrefatitle {The Holy Grail: A road map for unlocking the climate
  record stored within {M}ars’ polar layered deposits} {The holy grail: A
  road map for unlocking the climate record stored within {M}ars’ polar
  layered deposits}.{\BBCQ}
\newblock
\APACjournalVolNumPages{Planet. Space. Sci}{184}{}{104841}.
\newblock
\begin{APACrefDOI} \doi{https://doi.org/10.1016/j.pss.2020.104841}
  \end{APACrefDOI}
\PrintBackRefs{\CurrentBib}

\bibitem [\protect \citeauthoryear {%
Smith%
, Putzig%
, Holt%
\BCBL {}\ \BBA {} Phillips%
}{%
Smith%
\ \protect \BOthers {.}}{%
{\protect \APACyear {2016}}%
}]{%
smith2016}
\APACinsertmetastar {%
smith2016}%
\begin{APACrefauthors}%
Smith, I\BPBI B.%
, Putzig, N\BPBI E.%
, Holt, J\BPBI W.%
\BCBL {}\ \BBA {} Phillips, R\BPBI J.%
\end{APACrefauthors}%
\unskip\
\newblock
\APACrefYearMonthDay{2016}{}{}.
\newblock
{\BBOQ}\APACrefatitle {An ice age recorded in the polar deposits of {M}ars} {An
  ice age recorded in the polar deposits of {M}ars}.{\BBCQ}
\newblock
\APACjournalVolNumPages{Science}{352}{6289}{1075--1078}.
\newblock
\begin{APACrefDOI} \doi{10.1126/science.aad6968} \end{APACrefDOI}
\PrintBackRefs{\CurrentBib}

\bibitem [\protect \citeauthoryear {%
{Smith}%
}{%
{Smith}%
}{%
{\protect \APACyear {2004}}%
}]{%
smith2004}
\APACinsertmetastar {%
smith2004}%
\begin{APACrefauthors}%
{Smith}, M\BPBI D.%
\end{APACrefauthors}%
\unskip\
\newblock
\APACrefYearMonthDay{2004}{}{}.
\newblock
{\BBOQ}\APACrefatitle {Interannual variability in {TES} atmospheric
  observations of {Mars} during 1999–2003} {Interannual variability in {TES}
  atmospheric observations of {Mars} during 1999–2003}.{\BBCQ}
\newblock
\APACjournalVolNumPages{Icarus}{167}{1}{148--165}.
\newblock
\begin{APACrefDOI} \doi{0.1016/j.icarus.2003.09.010} \end{APACrefDOI}
\PrintBackRefs{\CurrentBib}

\bibitem [\protect \citeauthoryear {%
Sotzen%
\ \BBA {} Lewis%
}{%
Sotzen%
\ \BBA {} Lewis%
}{%
{\protect \APACyear {2020}}%
}]{%
stozen2020}
\APACinsertmetastar {%
stozen2020}%
\begin{APACrefauthors}%
Sotzen, J\BPBI P.%
\BCBT {}\ \BBA {} Lewis, K\BPBI W.%
\end{APACrefauthors}%
\unskip\
\newblock
\APACrefYearMonthDay{2020}{}{}.
\newblock
{\BBOQ}\APACrefatitle {The Significance of Sub-Milankovitch Signals in the
  Martian Northern Polar Layered Deposits} {The significance of
  sub-milankovitch signals in the martian northern polar layered
  deposits}.{\BBCQ}
\newblock
\APACjournalVolNumPages{Proc. Lunar Planet. Sci. Conf.}{}{}{}.
\newblock
\begin{APACrefDOI}
  \doi{https://ui.adsabs.harvard.edu/abs/2020LPI....51.2517S/abstract}
  \end{APACrefDOI}
\PrintBackRefs{\CurrentBib}

\bibitem [\protect \citeauthoryear {%
Villanueva%
\ \protect \BOthers {.}}{%
Villanueva%
\ \protect \BOthers {.}}{%
{\protect \APACyear {2015}}%
}]{%
villanueva2015}
\APACinsertmetastar {%
villanueva2015}%
\begin{APACrefauthors}%
Villanueva, G\BPBI L.%
, Mumma, M\BPBI J.%
, Novak, R\BPBI E.%
, K{\"a}ufl, H\BPBI U.%
, Hartogh, P.%
, Encrenaz, T.%
\BDBL {}Smith, M\BPBI D.%
\end{APACrefauthors}%
\unskip\
\newblock
\APACrefYearMonthDay{2015}{}{}.
\newblock
{\BBOQ}\APACrefatitle {Strong water isotopic anomalies in the martian
  atmosphere: Probing current and ancient reservoirs} {Strong water isotopic
  anomalies in the martian atmosphere: Probing current and ancient
  reservoirs}.{\BBCQ}
\newblock
\APACjournalVolNumPages{Science}{348}{6231}{218--221}.
\newblock
\begin{APACrefDOI} \doi{10.1126/science.aaa3630} \end{APACrefDOI}
\PrintBackRefs{\CurrentBib}

\bibitem [\protect \citeauthoryear {%
Vos%
}{%
Vos%
}{%
{\protect \APACyear {2022}}%
}]{%
vos2022dataset}
\APACinsertmetastar {%
vos2022dataset}%
\begin{APACrefauthors}%
Vos, E.%
\end{APACrefauthors}%
\unskip\
\newblock
\APACrefYearMonthDay{2022}{}{}.
\newblock
{\BBOQ}\APACrefatitle {Stratigraphic and Isotopic Evolution of the Martian
  Polar Caps from Paleo-Climate Models data set} {Stratigraphic and isotopic
  evolution of the martian polar caps from paleo-climate models data
  set}.{\BBCQ}
\newblock

\newblock
\begin{APACrefDOI} \doi{https://doi.org/10.5281/zenodo.5877622}
  \end{APACrefDOI}
\PrintBackRefs{\CurrentBib}

\bibitem [\protect \citeauthoryear {%
Vos%
, Aharonson%
\BCBL {}\ \BBA {} Schorghofer%
}{%
Vos%
\ \protect \BOthers {.}}{%
{\protect \APACyear {2019}}%
}]{%
vos2019}
\APACinsertmetastar {%
vos2019}%
\begin{APACrefauthors}%
Vos, E.%
, Aharonson, O.%
\BCBL {}\ \BBA {} Schorghofer, N.%
\end{APACrefauthors}%
\unskip\
\newblock
\APACrefYearMonthDay{2019}{}{}.
\newblock
{\BBOQ}\APACrefatitle {Dynamic and isotopic evolution of ice reservoirs on
  {M}ars} {Dynamic and isotopic evolution of ice reservoirs on {M}ars}.{\BBCQ}
\newblock
\APACjournalVolNumPages{Icarus}{324}{}{1--7}.
\newblock
\begin{APACrefDOI} \doi{10.1016/j.icarus.2019.01.018} \end{APACrefDOI}
\PrintBackRefs{\CurrentBib}

\bibitem [\protect \citeauthoryear {%
Webster%
\ \protect \BOthers {.}}{%
Webster%
\ \protect \BOthers {.}}{%
{\protect \APACyear {2013}}%
}]{%
webster2013}
\APACinsertmetastar {%
webster2013}%
\begin{APACrefauthors}%
Webster, C\BPBI R.%
, Mahaffy, P\BPBI R.%
, Flesch, G\BPBI J.%
, Niles, P\BPBI B.%
, Jones, J\BPBI H.%
, Leshin, L\BPBI A.%
\BDBL {}Steele, A.%
\end{APACrefauthors}%
\unskip\
\newblock
\APACrefYearMonthDay{2013}{}{}.
\newblock
{\BBOQ}\APACrefatitle {Isotope Ratios of {H}, {C}, and {O} in {CO2} and {H2O}
  of the Martian Atmosphere} {Isotope ratios of {H}, {C}, and {O} in {CO2} and
  {H2O} of the martian atmosphere}.{\BBCQ}
\newblock
\APACjournalVolNumPages{Science}{341}{6143}{260--263}.
\newblock
\begin{APACrefDOI} \doi{10.1126/science.1237961} \end{APACrefDOI}
\PrintBackRefs{\CurrentBib}

\end{thebibliography}
\end{document}